\documentclass[]{interact}

\usepackage{epstopdf}
\usepackage[caption=false]{subfig}
\usepackage{makecell}
\usepackage[T1]{fontenc}
\usepackage[english]{babel}		
\usepackage[utf8]{inputenc}
\usepackage{textcomp}
\usepackage{amsmath}
\usepackage{graphicx}
\usepackage{multirow}
\usepackage{geometry}
\usepackage{adjustbox}
\usepackage{xcolor}
\usepackage{lipsum}
\usepackage{subfig}
\usepackage{url}
\usepackage{caption}
\usepackage{etoolbox}
\usepackage{amsthm}
\usepackage{algorithm}
\usepackage[noend]{algpseudocode}
\usepackage{amssymb}
\usepackage{fancyhdr,lipsum}
\usepackage{lineno}

\usepackage[algo2e, ruled, lined, linesnumbered]{algorithm2e}
\usepackage[numbers,sort&compress]{natbib}
\bibpunct[, ]{[}{]}{,}{n}{,}{,}
\renewcommand\bibfont{\fontsize{10}{12}\selectfont}
\makeatletter
\def\NAT@def@citea{\def\@citea{\NAT@separator}}

\makeatother

\theoremstyle{plain}

\newcommand{\nonl}{\renewcommand{\nl}{\let\nl\oldnl}} 

\newtheorem*{remark}{Remark}

\theoremstyle{definition}

\theoremstyle{remark}

\begin{document}

\articletype{RESEARCH ARTICLE}

\title{Joint vehicle state and parameters estimation\\via Twin-in-the-Loop observers}

\author{
\name{F. Dettù\textsuperscript{a}\thanks{CONTACT F. Dettù. email: federico.dettu@polimi.it. S. Formentin, email: simone.formentin@polimi.it. S. M. Savaresi, email: sergio.savaresi@polimi.it.}. S. Formentin\textsuperscript{a} and S.~M. Savaresi\textsuperscript{a}}
\affil{\textsuperscript{a}Dipartimento di Elettronica, Informazione e Bioingegneria, Politecnico di Milano, Piazza Leonardo da Vinci 32, 20133, Milano, Italy.}
}

\maketitle

\begin{abstract}
Vehicular control systems are required to be both extremely reliable and robust to different environmental conditions, \emph{e.g.} load or tire-road friction. In this paper, we extend a new paradigm for state estimation, called Twin-in-the-Loop filtering (TiL-F), to the estimation of the unknown parameters describing the vehicle operating conditions. In such an approach, a digital-twin of the vehicle (usually already available to the car manufacturer) is employed on-board as a plant replica within a closed-loop scheme, and the observer gains are tuned purely from experimental data. The proposed approach is validated against experimental data, showing to significantly outperform the state-of-the-art solutions.
\end{abstract}


\section{Introduction}
\label{Section:Introduction}
The \textit{Twin-in-the-Loop} (TiL) framework has been recently proposed for vehicle dynamics control \citep{dettu_2022_SILC} and estimation of unknown states and variables \citep{riva_2022_SIL}. In this innovative framework, an high-fidelity off-the-shelf vehicle dynamics simulator is employed in a control/estimation loop - thanks to the recent developments in on-board hardware platforms \footnote{For example, \url{https://www.vi-grade.com/en/products/autohawk/}.}. Being the simulator a commercial product, we assume it to be a black-box, \emph{i.e.} we can only extract its outputs given some inputs, but we are not able to access its equations and update rules directly.
With regards to the TiL estimator, the authors in \citep{riva_2022_SIL} showed how the presented algorithm is able to estimate both unmeasured states (\emph{e.g.} vehicle sideslip) and other unknown variables, such as the tire-road forces. However, they did not address the estimation of time-varying parameters; in fact, modern vehicle dynamics controllers are strongly based on time-domain models of the process \citep{li_2016,theunissen_2020,tavernini_2020}. Due to the variability of conditions during the daily use of a vehicle - \emph{e.g.} the most common case being the payload - such controllers can underperform, or even destabilize the system, if such variations are not properly addressed. Hence, following a well established research trend on vehicle state and parameters estimation, we build this research within the TiL framework, extending it to a new application.

\textbf{Related works.} Existing literature solutions for four-wheeled vehicles on-line parameter estimation can be broadly split among direct sensing and software sensing ones. In the first class of solutions, a sensor for the measurement of the variable of interest is used (\emph{e.g.} \citep{yang_2008}, where the mass is estimated via strain gauges on the suspensions). The second class of solutions features instead the estimation of unmeasurable variables via production available sensors (\emph{i.e.} without adding new piece of hardware). If we focus on software sensing algorithms, we find many solutions in the literature - where we cite here the most relevant contributions in the past years - \citep{wenzel_2006,rozyn_2010,hong_2014,wielitzka_2015,reina_2017,zhu_2019,gong_2020,rodriguez_2021}, which can be categorized according to four axes: estimated quantities, employed sensors, used estimation algorithm, and employed model. Other important contributions design software sensing with an highly specific goal, like rollover risk evaluation \citep{wang_2020}: we do not consider them in this review.

On the estimated quantities side, the mass $M$ is indeed the most important vehicle parameter, and is always considered in the cited works; vehicle yaw inertia $J_{zz}$ is then considered in \citep{wenzel_2006,hong_2014,wielitzka_2015}, while roll and pitch inertia $J_{xx}$, $J_{yy}$ are considered in \citep{rozyn_2010}. Other estimated parameters include longitudinal \citep{wenzel_2006,rozyn_2010,hong_2014}, lateral \citep{rozyn_2010} or vertical \citep{wang_2020} center-of-mass position variation. The tire-road friction coefficient is estimated in \citep{rodriguez_2021} - however, this parameter is different in nature from the other considered in the literature, as its variations are not imputable to a variation of the vehicle load conditions, but rather to the terrain.

With regards to the sensing layout, one common denominator is the use of classical production vehicle sensors, like gyroscopes or Inertial Measurement Units (IMU); yaw rate, which is a gyroscope output, is employed in all the cited articles but \citep{rozyn_2010}. Acceleration information is used in all references; \citep{rozyn_2010} use three acceleration sensors, to be placed above the unsprung masses. Other considered sensors include wheel angular speed \citep{zhu_2019,rodriguez_2021}, suspension displacement \citep{rodriguez_2021}, GPS position measurements \citep{rodriguez_2021} and vehicle longitudinal speed \citep{wenzel_2006,wielitzka_2015,hong_2014}.

Moving further to the considered estimation algorithm, most works are based on Kalman Filtering (KF) \citep{wenzel_2006,wielitzka_2015,hong_2014,reina_2017,rodriguez_2021}, specifically declined - except for \citep{wielitzka_2015,reina_2017,rodriguez_2021} - in its \textit{Dual} formulation; \textit{Dual} KF is based on two separated observers, one devoted to classical state filtering, and the other one solely devoted to parameter estimation, to be switched-off once a good estimate of said parameters has been reached. \citep{zhu_2019} propose a similar formulation, where the correction law is computed via a particle filter. On the other hand, \citep{rozyn_2010} propose a different approach, based on the use of free decay responses of the vehicle - more similar to a classic identification algorithm rather than to a state observer.

Finally, when coming to employed plant replica, the simplest model is the single-track, which is however only used in \citep{reina_2017}, as it fails to capture most of vehicle dynamics non-linearities. The most common model is the double-track \citep{wenzel_2006,wielitzka_2015,hong_2014,reina_2017,zhu_2019}; the latter can be enhanced with roll dynamics, as in \citep{wenzel_2006,hong_2014}, or wheel dynamics \citep{zhu_2019}, depending on the estimation target or the available measurements. 
\citep{rozyn_2010} consider a full vehicle model - \emph{i.e.} featuring also unsprung mass dynamics, roll and pitch dynamics. Finally, the most complex model is the one proposed in \citep{rodriguez_2021}: the authors consider a multibody model - featuring 14 states. \\

\textbf{Contributions.} From the literature review, an important point emerges. As also pointed out in \citep{riva_2022_SIL}, classical vehicular state observers are based on simplified models of the variables to be estimated: an ad-hoc model for each variable is required, and each model has to be accurately calibrated. Given that approximately $75\%$ of the time of a control system design project is devoted to modelling, as indicated in \cite{gevers2005identification}, the resulting procedure turns out to be cumbersome.
Let this apart, consider that any estimator is based upon a set of hyperparameters controlling its behaviour - \emph{e.g.} the noise covariances in Kalman Filters. Generally speaking, the more complex an estimator is, the longer the hyperparameter selection is. Relying on simple correction laws is thus beneficial in the implementation phase.

In TiL framework we employ a single model - a digital twin - for each variable to be estimated: this solves the problem of designing ad-hoc models and of having many models running at run-time on the same estimation algorithm. The model calibration phase comes for free when using a digital twin, in the assumption that a car manufacturer already has a library of simulators for their production vehicles.
On the other hand, the correction algorithm complexity is now shifted to the model fidelity: the more faithful to the real system the model is, the less intense - and thus simple - the correction need to be. The authors in \citep{riva_2022_SIL} actually show as a linear and time-invariant correction law is sufficient to outperform the benchmark.

Matter-of-factly, the authors in \citep{rodriguez_2021} consider an high-fidelity model of the plant as in TiL framework: however, being the considered model a white-box, the authors are able to directly compute the update rules via a Kalman Filter-like algorithm. This is not possible in TiL, as the underlying model is a black-box, and the correction law, although linear, has to be tuned via model-free approaches \citep{riva_2022_SIL}.

With respect to the research in \citep{riva_2022_SIL}, we extend the TiL framework to the estimation of varying parameters, providing insights on how the correction law has to be designed, and on the feasibility of the approach in terms of robustness to measurement noise and uncertainty.\\

The paper is structured as follows. In Section \ref{Section:Problem_statement} we provide to the reader some preliminaries on the TiL estimator. Section \ref{Section:case_study} defines the problem of varying mass and moment of inertia, and its estimation. Then, in Section \ref{Section:simulation_analysis} we conduct an in-depth analysis of the parameters estimation problem by independently considering set of variables, providing sensitivity analyses to uncertainty and noise. Section \ref{Section:case_study_experimental} shows an application of the described and simulated framework on experimental data, while Section \ref{Section:Conclusions} concludes the manuscript with a few considerations.

\section{Twin-in-the-Loop observer architecture}\label{Section:Problem_statement}
\begin{figure}[h]
	\centering
	\includegraphics[width=0.8 \columnwidth]{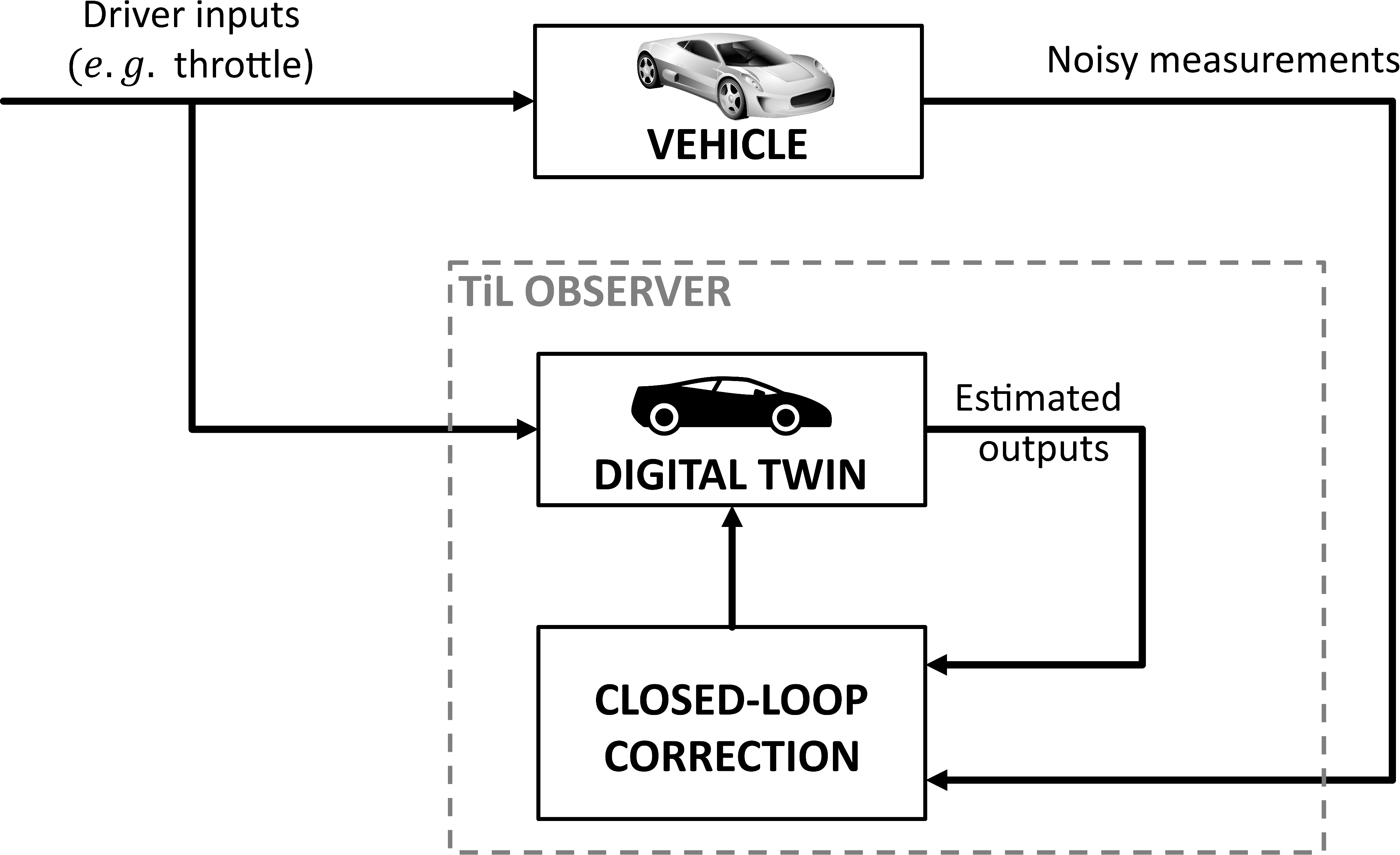}
	\caption{Twin-in-the-Loop estimation scheme.}
	\label{fig:estimator_til}
\end{figure}
\begin{figure*}[h]
	\centering
	\includegraphics[width=1 \columnwidth]{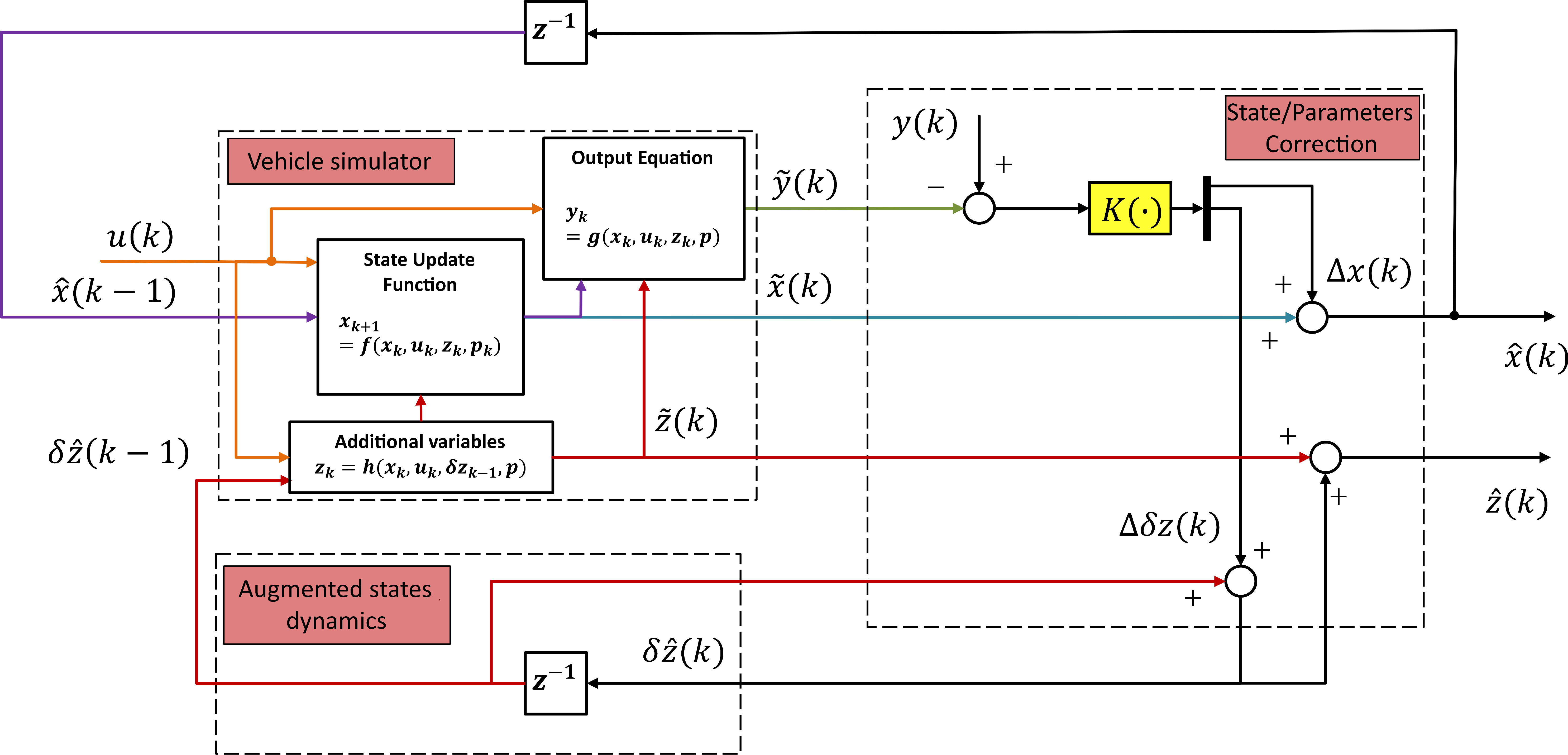}
	\caption{Twin-in-the-Loop complete estimation architecture, featuring correction of simulator states and augmented states.}
	\label{fig:estimator_architecture}
\end{figure*}
\citep{riva_2022_SIL} proposed an unified estimator for all the variables of interest in vehicle dynamics control. In such an architecture (see Fig. \ref{fig:estimator_til}), the two main features are the following
\begin{itemize}
	\item The vehicle simulator, an high-fidelity plant replica, whose equations are a black-box.
	\item The closed-loop correction, based upon available measurements, perturbs the model in order to properly estimate unknown variables.
\end{itemize} 

The detailed architecture of the complete observer is given in Fig. \ref{fig:estimator_architecture}.
Generically speaking, the simulator states at time $k$ are updated according to the following equations
\begin{equation}
	\begin{split}
		\tilde{x}_{k+1} &= f(\hat{x}_{k},u_{k}, \tilde{z}_{k},p), \\
			\tilde{y}_k & = g(\hat{x}_{k},u_k, 	\tilde{z}_k,p), \\
		\tilde{z}_k &= h(\hat{x}_k,u_k,\delta \hat{z}_{k-1},p)
	\end{split}
\end{equation}
where $f,\ g,\ h$ are unknown (possibly nonlinear) functions, $y \in \mathbb{R}^{n_y}$ is the set of measurable outputs, $x\in \mathbb{R}^{n_x}$ is the set of internal states of the simulator, $z\in \mathbb{R}^n{_z}$ is a set of additional variables to be estimated and $p$ is a set of constant parameters. We denote $\tilde{\nu}_k$ an a-priori estimation for a certain variable $\nu_k$: then closed-loop correction updates the a-priori estimates, and generates a-posteriori ones $\hat{\nu}_k$ - as in the widely known Kalman Filter formulation.

The authors in \citep{riva_2022_SIL} extend the state vector to estimate tire-road contact forces; more specifically, the additional variables $z$ are estimated by correcting the nominal values provided by the simulator
\begin{equation}
	\hat{z}_k=\tilde{z}_k+\delta \hat{z}_k.
\end{equation}
The artificial variable $\delta \hat{z}_k$ is described by a fictitious constant dynamics equation
\begin{equation}
	\delta \tilde{z}_{k+1}=\delta \hat{z}_k.
\end{equation}
The latter assumption is widely used in case the parameters to be estimated are slowly varying.
Overall, the set of states, including internal and extended ones, now reads $x^{aug}=\begin{bmatrix}
	x^T &\delta z^T
\end{bmatrix}$.
At this point, the innovation term at step $k$, consisting in the mismatch among measured ($y_k$) and a-priori estimated outputs ($\tilde{y}_k$) is to be used to correct the state vector $x^{aug}$. The innovation is mapped onto the states via a linear law in \citep{riva_2022_SIL}
\begin{equation}
	\begin{split}
		\Delta x_k^{aug}&=K\left(y_k-\tilde{y}_k\right) \\
		\hat{x}^{aug}_k&=\tilde{x}^{aug}_k+\Delta x^{aug}_k.
		\label{eq:correction_linear}
	\end{split}
\end{equation}
However, we will show in the following that the linear formulation is inadequate for the estimation of certain parameters, \emph{e.g.} the vehicle mass or inertia. Let us thus define a mixed linear-nonlinear correction law
\begin{equation}
	\label{eq:correction_nonlinear}
	\hat{x}_k^{aug}=\tilde{x}_k^{aug}+\begin{bmatrix}
		K & 0_{} \\
		0 & \mathcal{K}\left(y_k\right)
	\end{bmatrix}\left(y_k-\tilde{y}_k\right),
\end{equation}
where $K\in \mathbb{R}^{n_x\times n_y}$ maps the innovation onto the nominal states, and $\mathcal{K}\left(y_k\right)\in \mathbb{R}^{n_z\times n_y}$ maps the innovation onto the extended states.

\section{A parameter estimation case study: the uncertain load setting}
\label{Section:case_study}
\begin{figure}[h]
	\centering
	\includegraphics[width=0.7 \columnwidth]{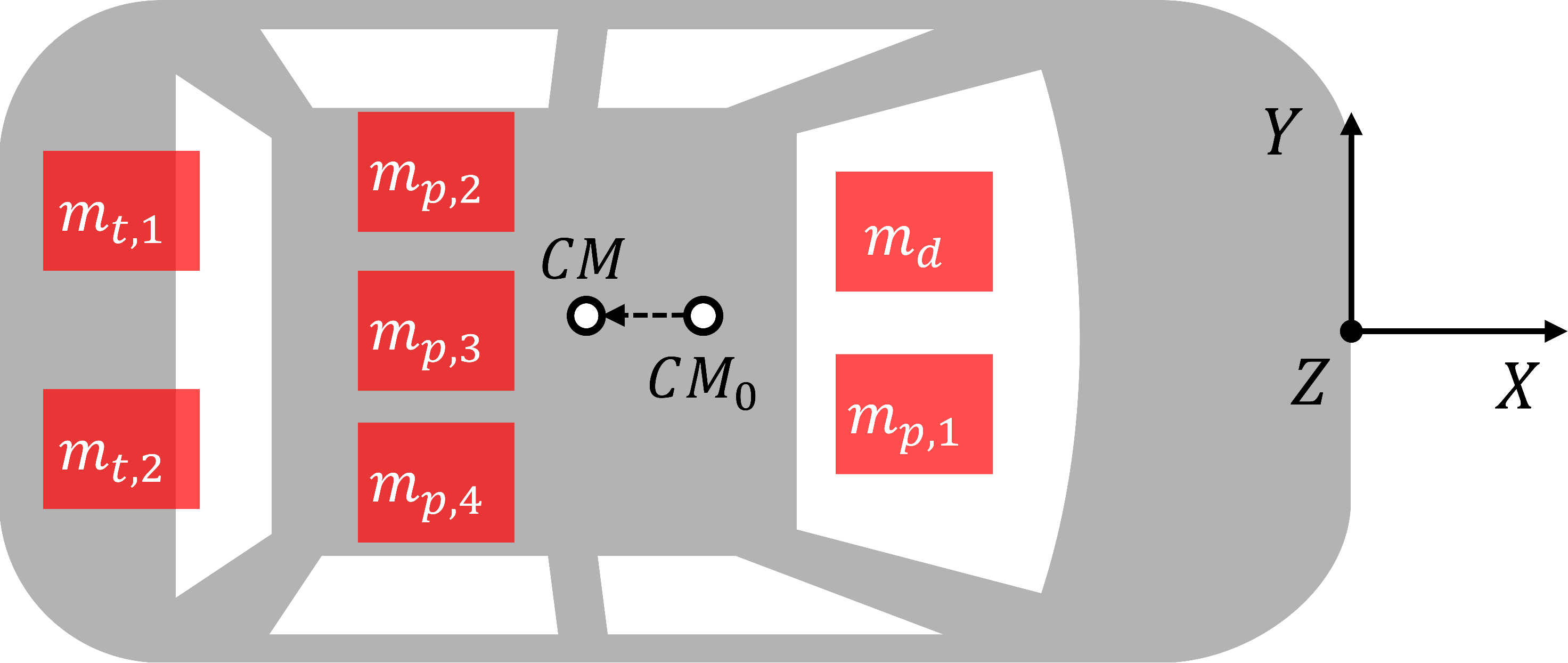}
	\caption{Vehicle top view, considering differently added additional masses.}
	\label{fig:vehicle_layout}
\end{figure}
Among the physical parameters defining a vehicle, mass and inertia might significantly change due to different load conditions, while being extremely important in several on-board control systems. Hence, we consider in the following the problem of estimating such parameters in a road vehicle via the TiL estimator architecture.
Consider the 5-seats vehicle in Fig. \ref{fig:vehicle_layout}. Such vehicle can be loaded with additional masses. On the mathematical perspective, we assume each of the $m_i,\ i=1,\ldots,\mathcal{M}$ to be a point mass, characterized by a certain position $\textrm{p}_{m_i}=\left(p_{m_{i,x}},p_{m_{i,y}},p_{m_{i,z}}\right)$ in the reference frame defined by axes $X,\ Y,\ Z$ - as displayed in Fig. \ref{fig:vehicle_layout}. If we assume that the vehicle chassis can be modeled as a rigid body with its mass concentrated in a single point - a common assumption in vehicle dynamics - the center-of-mass of such system is computed as \citep{thornton_2004}
\begin{equation}
	\begin{split}
		CM_x&=\sum_{i=0,\ldots,\mathcal{M}} m_i\cdot p_{m_{i,x}}, \\
		CM_y&=\sum_{i=0,\ldots,\mathcal{M}} m_i\cdot p_{m_{i,y}}, \\
		CM_z&=\sum_{i=0,\ldots,\mathcal{M}} m_i\cdot p_{m_{i,z}}.
	\end{split}
\end{equation}
whereas $m_0$ denotes the original lumped mass of the chassis.
Indeed, the center-of-mass including additional masses - $CM=\left(d_x,d_y,d_z\right)$ - is going to be different from the original one - $CM_0=\left(d_{0,x},d_{0,y},d_{0,z}\right)$. The new lumped mass is then simply defined as 
\begin{equation}
	M_{tot} = m_0 + \sum_{i=1,\ldots,\mathcal{M}} m_i.
\end{equation}
Consequently, the chassis moments and products of inertia change due to the mass variation and the re-defined center-of-mass \citep{thornton_2004}. Considering the Huygens-Steiner theorem \cite{gobbi_2011}, we obtain the new moments/products of inertia, defined with respect to $CM$; this is done starting from the ones defined in $CM_0$ and taking into account the newly added masses. \emph{E.g.}, for the moments of inertia - $J_{xx},\ J_{yy},\ J_{zz}$ - one has
\begin{equation}
	\begin{split}
		J_{xx}=&J_{xx,0}-m_0\left[\delta\left(d_y,d_{y,0}\right)+\delta\left(d_z,d_{z,0}\right)\right] + \\
		&\sum_{i=1,\ldots,\mathcal{M}}m_i\left[\delta\left(p_{m_i,y}, d_y\right)+\delta\left(p_{m_i,z}, d_z\right)\right], \\
		J_{yy}=&J_{yy,0}-m_0\left[\delta\left(d_x,d_{x,0}\right)+\delta\left(d_z,d_{z,0}\right)\right] + \\
		&\sum_{i=1,\ldots,\mathcal{M}}m_i\left[\delta\left(p_{m_i,x}, d_x\right)+\delta\left(p_{m_i,z}, d_z\right)\right], \\
		J_{zz}=&J_{zz,0}-m_0\left[\delta\left(d_x,d_{x,0}\right)+\delta\left(d_y,d_{y,0}\right)\right] + \\
		&\sum_{i=1,\ldots,\mathcal{M}}m_i\left[\delta\left(p_{m_i,x}, d_x\right)+\delta\left(p_{m_i,y}, d_y\right)\right], \\
	\end{split}
\end{equation}
whereas $J_{xx,0},\ J_{yy,0},\ J_{zz,0}$ represent the chassis nominal moment of inertia, and $\delta\left(x_1,x_2\right)=\left(x_1-x_2\right)^2$. Also product of inertia variations - \emph{e.g.} $J_{xy}$- can be quantified analytically in a similar way; for the load conditions considered in the case study - realistic ones for road vehicles - we don't have significant variations of these parameters.

Now, let us consider again scheme of Fig. \ref{fig:estimator_til}: the high-fidelity digital twin is modelled with a set of nominal parameters, \emph{i.e.} lumped mass, inertia, and center-of-mass location. The real vehicle is instead loaded with additional masses, and thus ''perturbed'' from the nominal parameters set.

In \citep{riva_2022_SIL}, the authors design a closed-loop correction so as to force the simulator states to be identical to the ones of the real vehicle, regardless of the reasons leading to differences between the two instances of the vehicle. On the other hand, in the present work we shift the philosophy, trying to estimate also the exact differences among the two instances.
\section{Simulation analysis}
\label{Section:simulation_analysis}
\begin{table}[h]
	\centering
	\begin{minipage}[b]{0.49\columnwidth}
		\centering
		\begin{tabular}{|l|ll|}
			\hline
			\textbf{Nominal param.} & \multicolumn{2}{c|}{\textbf{Value}}         \\ \hline
			$m_{0}\ [kg]$                             & \multicolumn{2}{l|}{$2125.8$}          \\ \hline
			$J_{xx,0}\ [kg\cdot m^2]$                          & \multicolumn{2}{l|}{$834.23$}  \\ \hline
			$J_{yy,0}\ [kg\cdot m^2]$           & \multicolumn{2}{l|}{$3640.182$} \\ \hline
			$J_{zz,0}\ [kg\cdot m^2]$                          & \multicolumn{2}{l|}{$3932.77$} \\ \hline
			$J_{xy,0}\ [kg\cdot m^2]$                          & \multicolumn{2}{l|}{$0.14$}    \\ \hline
			$J_{xz,0}\ [kg\cdot m^2]$                          & \multicolumn{2}{l|}{$0.097$}   \\ \hline
			$J_{yz,0}\ [kg\cdot m^2]$                          & \multicolumn{2}{l|}{$3.86$}    \\ \hline
			$d_{0,x}\ [cm]$                           & \multicolumn{2}{l|}{$-125.0$}             \\ \hline
			$d_{0,y}\ [cm]$                           & \multicolumn{2}{l|}{$-0.003$}                 \\ \hline
			$d_{0,z}\ [cm]$                           & \multicolumn{2}{l|}{$64.4$}             \\ \hline
		\end{tabular}
		\caption{Nominal vehicle mass, inertia and CM parameters.}
		\label{tab:nominal_table}
	\end{minipage}
	\hfill
	\begin{minipage}[b]{0.49\columnwidth}
		\centering
		\begin{tabular}{|l|ll|}
			\hline
			\textbf{Additional load param.} & \multicolumn{2}{c|}{\textbf{Value}} \\ \hline
			$m_{p,1}\ [kg]$                     & \multicolumn{2}{l|}{$75\ kg$}       \\ \hline
			$m_{p,2}\ [kg]$                     & \multicolumn{2}{l|}{$80\ kg$}       \\ \hline
			$m_{p,3}\ [kg]$                     & \multicolumn{2}{l|}{$65\ kg$}       \\ \hline
			$m_{p,4}\ [kg]$                     & \multicolumn{2}{l|}{$75\ kg$}       \\ \hline
			$m_{t,1}\ [kg]$                     & \multicolumn{2}{l|}{$30\ kg$}       \\ \hline
			$m_{t,2}\ [kg]$                     & \multicolumn{2}{l|}{$30\ kg$}       \\ \hline
		\end{tabular}
		\caption{Additional load parameters.}
		\label{tab:additional_table}
	\end{minipage}
\end{table}

\begin{table}[]
	\centering
	\begin{tabular}{|l|l|l|}
		\hline
		\textbf{Perturbed param.} & \multicolumn{1}{c|}{\textbf{Value}} & \textbf{Variation\footnote{Where the variation from value $v_0$ to value $v_1$ is defined as $\Delta=100(v_0/v_1-1)$.}} \\ \hline
		$M_{tot}\ [kg]$                   & $2480.8$                            & $+16.70\ \%$                     \\ \hline
		$J_{xx}\ [kg\cdot m^2]$               & $901.9$                            & $+8.11\ \%$                      \\ \hline
		$J_{yy}\ [kg\cdot m^2]$               & $4394.4$                            & $+20.72\ \%$                     \\ \hline
		$J_{zz}\ [kg\cdot m^2]$               & $4760.0$                            & $+21.03\ \%$                     \\ \hline
		$d_x\ [cm]$               & $-131.6$                            & $+5\ \%$                      \\ \hline
		$d_y\ [cm]$               & $1.6$                            & $+100.2\ \%$                     \\ \hline
		$d_z\ [cm]$               & $68.4$                            & $+5.3\ \%$                     \\ \hline
	\end{tabular}
	\caption{Perturbed vehicle parameters.}
	\label{tab:perturbed_table}
\end{table}
In this Section, we demonstrate the TiL methodology for the case study of Section \ref{Section:case_study}. For a better assessment of the potentialities and limitations of the present method, we conduct a preliminary analysis in a controlled simulation environment.
The nominal chassis parameters of the considered vehicle are given in Table \ref{tab:nominal_table}. Said nominal parameters also include the presence of a driver ($m_d$ in Fig. \ref{fig:vehicle_layout}). The parameters are those of a generic sport utility vehicle, which is modeled in VI-CarRealTime (CRT) simulation environment \citep{vigrade_2022}. The nominal vehicle model is perturbed by adding masses according to the scheme of Fig. \ref{fig:vehicle_layout}; the values of such loads are given in Table \ref{tab:additional_table}. For this new configuration, the center-of-mass, mass and inertia can be analytically computed according to what described in Section \ref{Section:case_study}; the new values for the most relevant parameters are in Table \ref{tab:perturbed_table}.

We apply the TIL architecture to the problem herein, defining the complete extended state vector as 
\begin{equation}
	x^{aug}= \begin{bmatrix}
		x^T & \delta M & \delta J_{xx} & \delta J_{yy} & \delta J_{zz}
	\end{bmatrix}^T,
\end{equation}
where $x$ is the set of VI-CarRealTime states, $x\in R$.
In this simulation analysis, only extended states - parameters - are going to be corrected, as we want to focus on the feasibility of parameters correction. Each extended state represents the deviation of said parameter from the nominal value provided by the simulator as an output. Joint state and parameter correction will be considered in an experimental case study in Section \ref{Section:case_study_experimental}.

In order to reproduce a simulation environment the closer to a real scenario, we introduce measurement noise. The noise is introduced on the signals coming from an - simulated in CRT - inertial measurement unit (IMU) placed near the vehicle CM. The following noise law is considered
\begin{equation}
	s_n=s+n_s,\ n_s\sim  WN(0,\sigma_s^2),
\end{equation}
where $s$ is a signal coming from an IMU, and $\sigma_s$ is the standard deviation of a suitable white noise. In the simulation case study, $x$-axis acceleration $a_x$ and $x,\ y,\ z$ axes angular rates $\omega_{x,y,z}$ are employed as measurements in the observer, hence, only these signals are perturbed with noise. When not specified, the $snr$ is tuned to be $\approx 10$: however, we also conduct sensitivity analyses to verify the impact of measurement noise onto the estimator.

When quantifying the observer performance, we consider as a metric the root mean square estimation error in the last $w=1$ seconds of test, as the dynamics of the estimated parameters are rather slow, and considering larger windows would take into account transient effects.
\begin{equation}
	rms_v= \dfrac{1}{wf_s}\sum_{k=N_s-wf_s}^{N_s} \left(\nu_k-\hat{\nu}_k\right)^2,
\end{equation}
Where $N_s$ is the number of samples in an experiment, $f_s\ [Hz]$ is the sampling frequency, and $\nu$ is a suitable parameter or variable to be estimated, \emph{e.g.} $\nu=\delta M$.

On the other hand, we also consider the percentage version of $rms_w$, in order to quantify the amount of error within the parameters values. This metric is defined as
\begin{equation}
	rms_{\nu,\%}=rms_\nu / \nu_0,
\end{equation}
Where $\nu_0$ is the nominal value of a variable to be estimated; \emph{e.g.} in our case, the nominal value for $\delta M$ is the difference among the perturbed value in Table \ref{tab:perturbed_table} and the nominal value in Table \ref{tab:nominal_table}.

In the following, we consider as acceptable a percentage $rms$ smaller than the $10\ \%$ of a perturbated variable.
\subsection{Center-of-mass position identification}
The $CM$ position is perturbed from $CM_0$ in the modified layout, as evident from Table \ref{tab:perturbed_table}. Most significant displacement occur along $x$ and $z$ axes, as $y$ axis variation are mostly due to extremely unbalanced - on the left-right direction - loads, which are not common in road vehicles. Given that the digital twin computes the torque and force balances referring to the center-of-mass, a not exact knowledge of $CM$ might be detrimental for estimating the other parameters.

Two possible ways of solving this problem exist:
\begin{itemize}
	\item Insert the variables $d_x,\ d_y,\ d_z$ onto the augmented state vector $x^{aug}$, and consequently correct the mismatch from the nominal values;
	\item Identify in a traditional way - see \emph{e.g.} \citep{selim_2008,huang_2014} - $d_x,\ d_y,\ d_z$ on-line via available data and simplified models.
\end{itemize}
In the present work, we decide to follow the second approach, as we want to concentrate onto the estimation of mass and inertia: hence, we assume correct knowledge of $CM$. However, we also perform sensitivity analyses to check the effect of a wrong $CM$ onto the estimation performance: in general, we will show as for the considered load conditions, which are reasonable for a passenger car, the performance loss is negligible even in case of wrong $CM$ estimation.
\subsection{Vehicle mass estimation}
\label{section:vehicle_mass_estimation}
\begin{figure}[h]
	\centering
	\subfloat[Mass estimation in an urban driving context.\label{fig:mass_est}]{\includegraphics[width=0.49 \columnwidth]{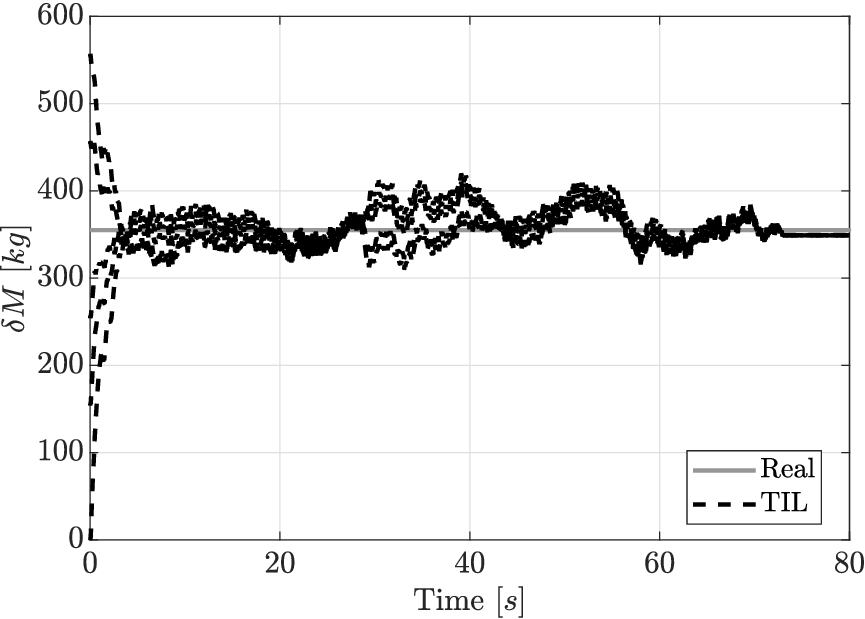}}
	\hfill
	\subfloat[Urban driving commands.\label{fig:mass_est_commands}]{\includegraphics[width=0.49 \columnwidth]{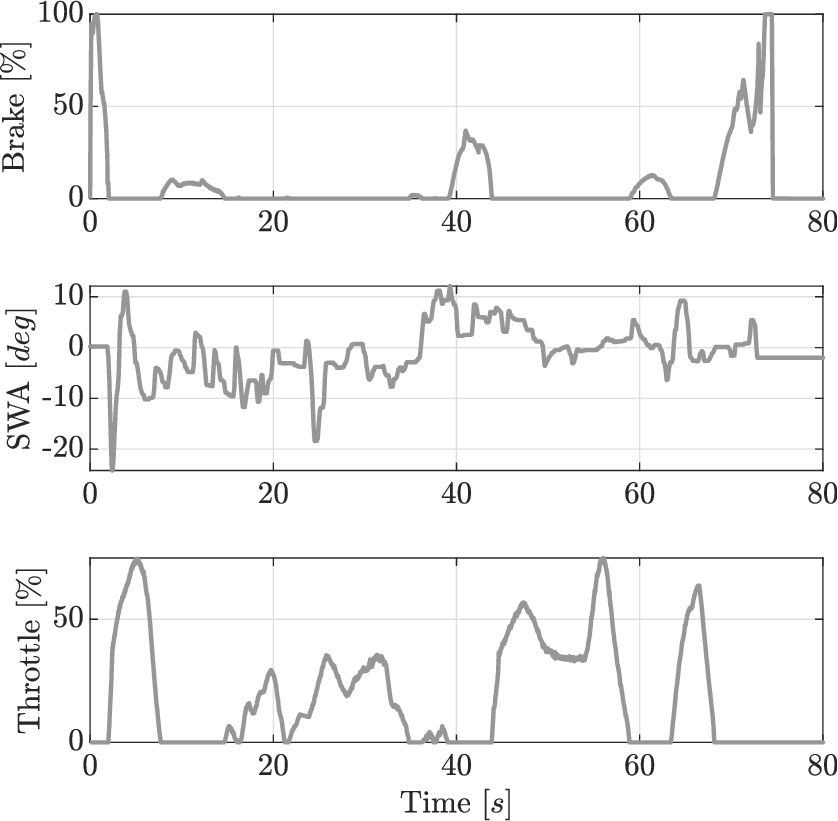}}
	\caption{Mass estimation in an urban driving like context, with noisy measurements. Estimated mass for different initial conditions.}
\end{figure}

\begin{figure}[h]
	\centering
	\begin{minipage}{0.49\textwidth}
		\centering
		\includegraphics[width=\textwidth]{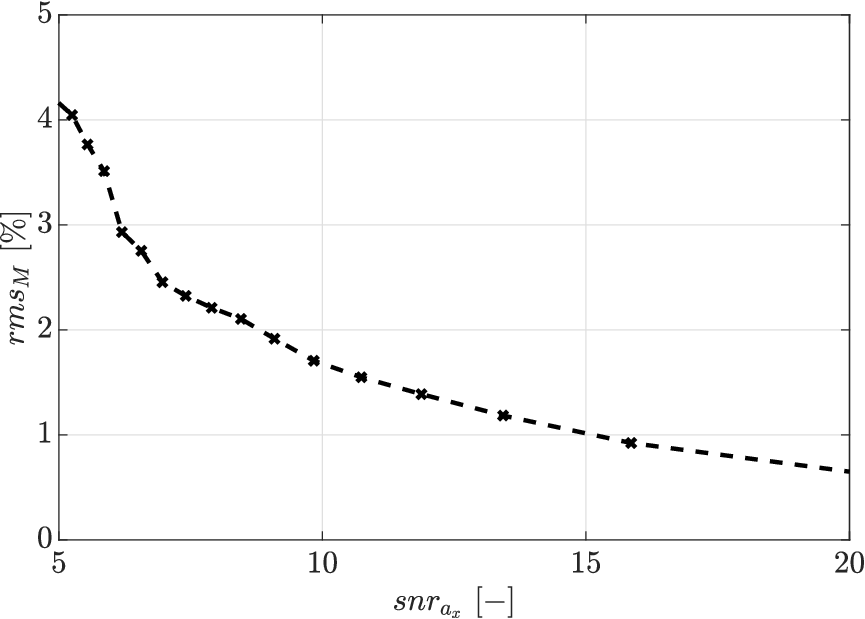}
		\caption{Mass estimation with varying noise levels.}
		\label{fig:noise_sensitivity_mass}
	\end{minipage}
	\hfill
	\begin{minipage}{0.49\textwidth}
		\centering
		\begin{tabular}{|l|l|} 
			\hline
			\textbf{Test conditions}                                                                                                               & \boldmath{$rms_{M}\ [\%]$}  \\ 
			\hline
			\begin{tabular}{@{\labelitemi\hspace{\dimexpr\labelsep+0.5\tabcolsep}}l@{}}Noiseless data ($snr=\infty$);\\Correct $CM$\end{tabular} & $0.64$            \\ 
			\hline
			\begin{tabular}{@{\labelitemi\hspace{\dimexpr\labelsep+0.5\tabcolsep}}l@{}}Noisy data ($snr=10$);\\Correct $CM$.\end{tabular}        & $1.61$              \\ 
			\hline
			\begin{tabular}{@{\labelitemi\hspace{\dimexpr\labelsep+0.5\tabcolsep}}l@{}}Noisy data ($snr=10$);\\Wrong $CM$.\end{tabular}          & $0.68$             \\
			\hline
		\end{tabular}
		\captionsetup{type=table}
		\caption{Mass estimation performance in different conditions.}
		\label{tab:mass_table}
	\end{minipage}
\end{figure}

Vehicle chassis mass can be estimated by suitably exciting the longitudinal dynamics. Consider the following simplified longitudinal dynamics balance for a road vehicle
\begin{equation}
	M_{chassis}\cdot a_x=F_{trac}-F_{brake}-F_{roll}-F_{aero}.
	\label{eq:longitudinal_dynamics}
\end{equation}
It is clear that, provided that roughly the same forces act on the vehicle, an increased mass yields a lower longitudinal acceleration. Hence, if the digital twin is characterized by a smaller mass than the real vehicle, a suitable correction can be based upon the difference between longitudinal accelerations. Indeed, if the terms on the right hand side of Eq. \eqref{eq:longitudinal_dynamics} are significantly different, we are not able to determine whether an acceleration variation is due to a mass difference or rather to an error in modelling the brake system or the tire-road interaction. For this reason, in this simulation analysis the sole modelling difference among the two simulated instances of the vehicle lie in the mass perturbation. In Section \ref{Section:case_study_experimental} we will test the mass estimation on real data, where other modelling differences might exist.

The correction applied on $\delta M$ at time step $k$ reads
\begin{equation}
	\Delta \delta M_k=K_{a_{x}-\delta M}\cdot  \textrm{sgn}(a_{x,k})\cdot (a_{x,k}- \tilde{a}_{x,k}),
	\label{eq:mass_correction}
\end{equation}
Where $K_{a_x,\delta M}$ is a suitable gain to be tuned and $\textrm{sgn}\left(\cdot\right)$ extracts the sign of the argument. Note as $\mathcal{K}\left(y_k\right)=K_{a_{x}-\delta M}\cdot  \textrm{sgn}(a_{x,k})$ in this context.

The importance of the sign operator is easily understood with a simple example. Consider a coasting down - deceleration due to inertia $a_x<0$ - event: if the vehicle has an higher mass than the simulator, it will coast down faster, thus $\left(a_x - \tilde{a}_x\right)<0$. Correcting $\delta M$ proportionally to $\left(a_x - \tilde{a}_x\right)$, with the same sign, would lead to a decrement of the extended state: however, $\delta M$ should instead increase to compensate for the greater mass on the vehicle. Thus, the introduction of the sign operator accounts for this issue, an introduces a necessary non-linearity in the correction law.

The considered experiment for the mass estimation is an urban driving one - see Fig. \ref{fig:mass_est_commands}: in said test, only the correction of Eq. \eqref{eq:mass_correction} is applied.

The results - in case of noisy data and correct center-of-mass - can be appreciated in Fig. \ref{fig:mass_est}: the mass is correctly estimated even by starting from different initial conditions, and the estimate eventually converges to the correct value. 
Table \ref{tab:mass_table} reports the estimation performance in different conditions: as one can notice, the mass estimate is highly robust both to measurement noise and to imperfect knowledge of the center-of-mass. In the worst case, $rms_M=0.68\ \%$: since $\delta M=355\ kg$, the error is smaller than $3\ kg$.

Figure \ref{fig:noise_sensitivity_mass} compares the estimator performance for an $snr$ growing from $5$ to $20$: we can conclude that for reasonable noise levels, the estimator performance does not drop below the $10 \%$, which was originally set to be our limit.

Finally, note that $\delta M$ converges to the real value even if we are not correcting $\delta J_{xx},\ \delta J_{yy}, \delta J_{zz}$: this highlights the good decoupling among the estimated variables, without the need for designing a separated model for each of them.

\subsection{Yaw inertia and roll inertia estimation}
\begin{figure}[h]
	\centering
	\subfloat[Correct center-of-mass estimation case.\label{fig:roll_yaw_inertia_est}]{\includegraphics[width=0.48\columnwidth]{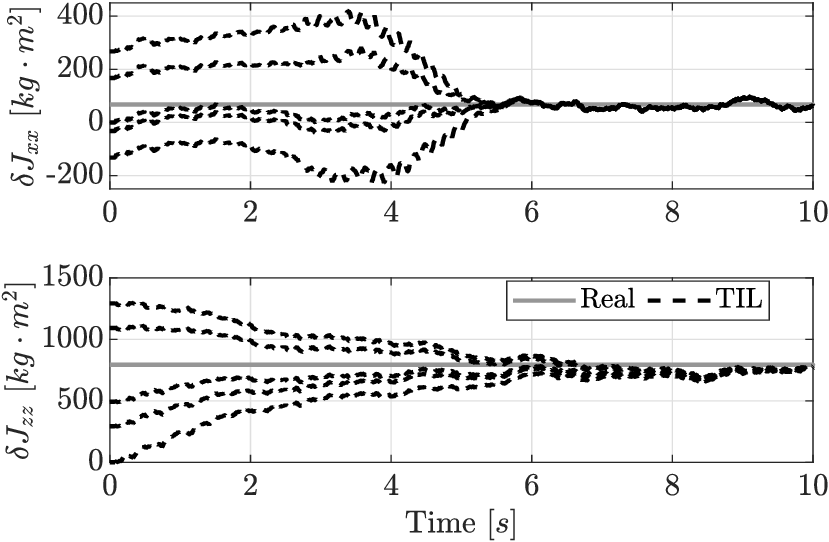}}
	\hfill
	\subfloat[Wrong center-of-mass estimation case.\label{fig:roll_yaw_inertia_withMass_wrongCM_est}]{\includegraphics[width=0.48\columnwidth]{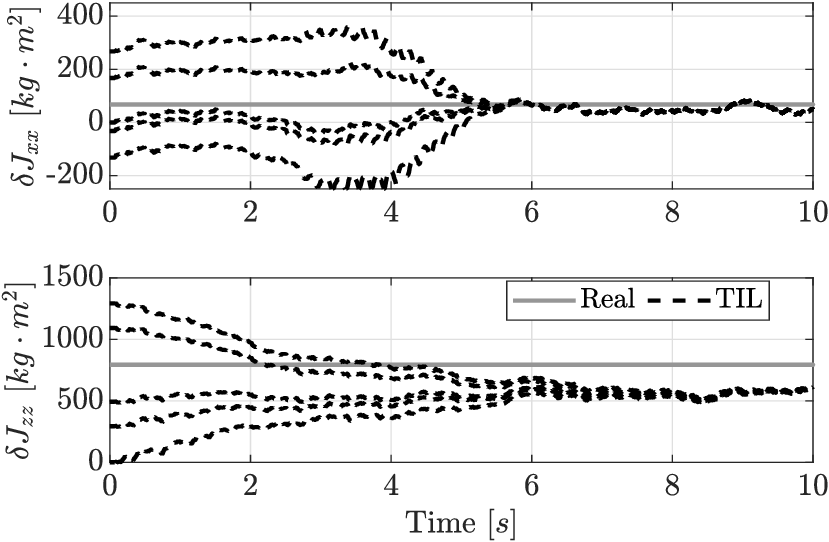}}
	\caption{Roll and yaw inertia estimation by means of a swept steer experiment, with noisy measurements. The upper plot depicts the estimated parameters in case the center-of-mass is exactly known, while the lower one depicts them in case the CM is not exactly known.}
	\label{fig:roll_yaw_inertia_estimation}
\end{figure}
%

\begin{figure}[h]
	\centering
	\begin{minipage}[b]{0.49\columnwidth}
		\centering
		\includegraphics[width=\textwidth]{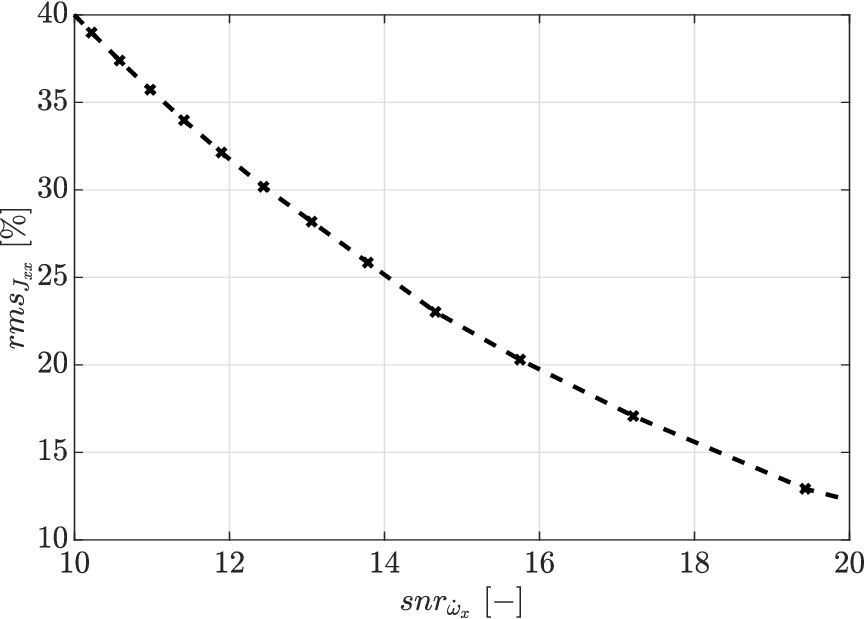}
		\caption{Roll inertia estimation with varying noise levels.}
		\label{fig:noise_sensitivity_Jxx}
	\end{minipage}
	\hfill
	\begin{minipage}[b]{0.49\columnwidth}
		\centering
		\includegraphics[width=\textwidth]{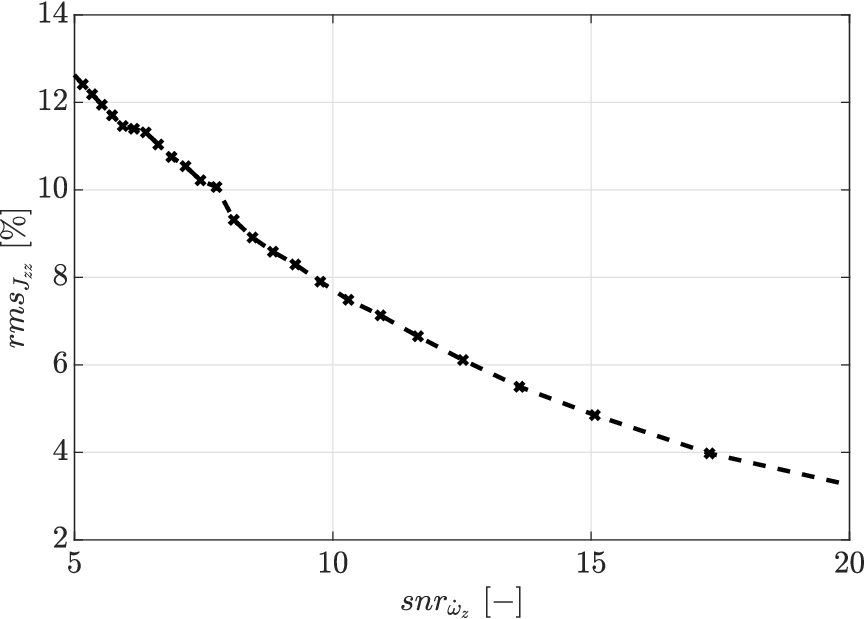}
		\caption{Yaw inertia estimation with varying noise levels.}
		\label{fig:noise_sensitivity_Jzz}
	\end{minipage}
\end{figure}

Roll and yaw inertia can be estimated by exciting lateral dynamics and left-to-right load transfer, which often come together, \emph{e.g.} while negotiating a curve. A proper experiment to estimate both variables at the same time is a swept steer maneuver, at constant speed, from $0$ to $4$ $Hz$. The non-linear correction law to be applied onto $\delta J_{xx}$ and $\delta J_{zz}$ follows from what described in Section \ref{section:vehicle_mass_estimation} for the mass 
\begin{equation}
	\begin{split}
		\Delta \delta J_{xx,k_k}&=K_{\dot{\omega}_{x}-\delta J_{xx}}(\dot{\omega}_{x,k}- \tilde{\dot{\omega}}_{x,k})\cdot \textrm{sgn}(\dot{\omega}_{x,k}), \\	
		\Delta \delta J_{zz,k_k}&=K_{\dot{\omega}_{z}-\delta J_{xx}}(\dot{\omega}_{z,k}- \tilde{\dot{\omega}}_{z,k})\cdot \textrm{sgn}(\dot{\omega}_{z,k}).
	\end{split}
	\label{eq:Jxx_Jzz_correction}
\end{equation}
Where the angular accelerations $\dot{\omega}$ are employed - such signals are easily obtained from the angular rates via differentiation.
Note that while estimating $J_{xx}$ and $J_{zz}$, $\delta M$ could have been already estimated as described before. Same goes for the center-of-mass.
Figure \ref{fig:roll_yaw_inertia_est} depicts the estimation results in case of noisy measurements, wrong mass estimate, and correct center-of-mass; only corrections in Eq. \eqref{eq:Jxx_Jzz_correction} are applied. The estimates converge to the true value, even starting from different initial conditions. On the other hand, Fig. \ref{fig:roll_yaw_inertia_withMass_wrongCM_est} depicts the estimation of the same variable, in case the nominal center-of-mass is kept within the digital twin: as expected, we note a steady-state error in the estimated inertia - mostly noticeable for $J_{zz}$.

Table \ref{tab:roll_yaw_table} quantifies the estimation performance in different cases. One can notice as the error on $J_{xx}$ is generally higher and grows faster as the uncertainty increases. This is easily explained as the nominal value and the variation of $J_{xx}$ are much smaller than for the other inertia. Finally, Fig. \ref{fig:noise_sensitivity_Jxx} and Fig. \ref{fig:noise_sensitivity_Jzz} show the estimator performance when considering different noise levels on the employed measurements. As expected, $J_{xx}$ estimation performance significantly drops, and is never within the $10\%$ limit. On the other hand, $J_{zz}$ estimation shows good performance up to $snr\approx 7.5$, which is a reasonable bound for yaw acceleration noise.

\begin{table*}
	\centering
	\caption{Roll and yaw inertia estimation performance in different conditions.}
	\label{tab:roll_yaw_table}
	\begin{tabular}{|l|l|l|} 
		\hline
		\textbf{Test conditions}                                                                           & \boldmath{$rms_{J_{xx}}\ [\%]$} & \boldmath{$rms_{J_{zz}}\ [\%]$}  \\ 
		\hline
		\begin{tabular}[c]{@{}l@{}}Noiseless data ($snr=\infty$);\\Correct mass;\\Correct CM.\end{tabular} & $0.38$           & $0.96$            \\ 
		\hline
		\begin{tabular}[c]{@{}l@{}}Noiseless data ($snr=\infty$);\\Wrong mass;\\Correct CM.\end{tabular}   & $4.22$           & $0.25$            \\ 
		\hline
		\begin{tabular}[c]{@{}l@{}}Noisy data ($snr=10$);\\Correct mass;\\Correct CM.\end{tabular}         & $23.91$          & $6.18$            \\ 
		\hline
		\begin{tabular}[c]{@{}l@{}}Noisy data ($snr=10$);\\Wrong mass;\\Correct CM.\end{tabular}         & $26.50$          & $5.38$            \\ 
		\hline
		\begin{tabular}[c]{@{}l@{}}Noisy data ($snr=10$);\\Wrong mass;\\Wrong CM.\end{tabular}             & $45.43$          & $26.50$           \\
		\hline
	\end{tabular}
\end{table*}

\subsection{Pitch inertia estimation}
\begin{figure}[th]
	\centering
	\begin{minipage}[b]{0.49\columnwidth}
		\centering
		\includegraphics[width=\textwidth]{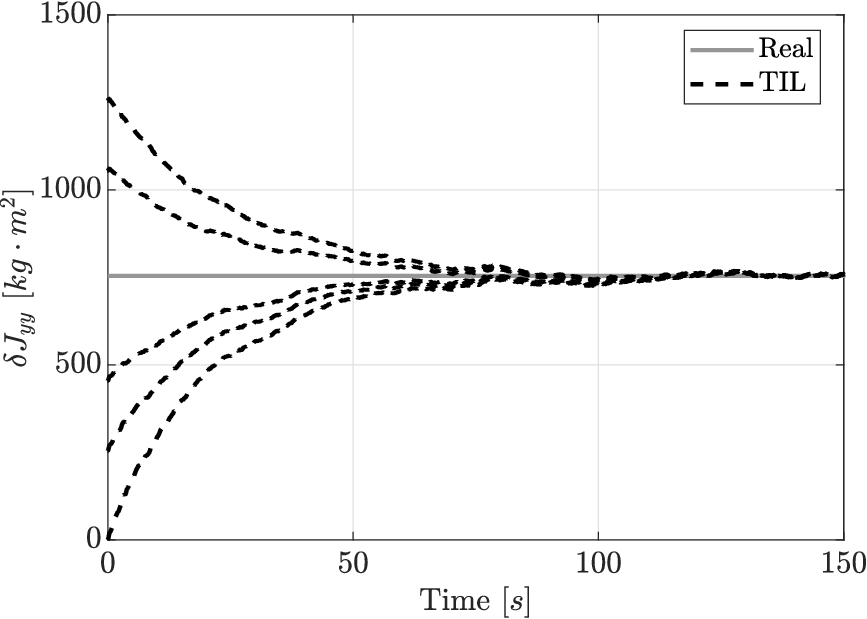}
		\caption{Pitch inertia estimation in case the road is known without uncertainty, with noisy measurements.}
		\label{fig:pitch_inertia_withMass_roadUncert_est}
	\end{minipage}
	\hfill
	\begin{minipage}[b]{0.49\columnwidth}
		\centering
		\includegraphics[width=\textwidth]{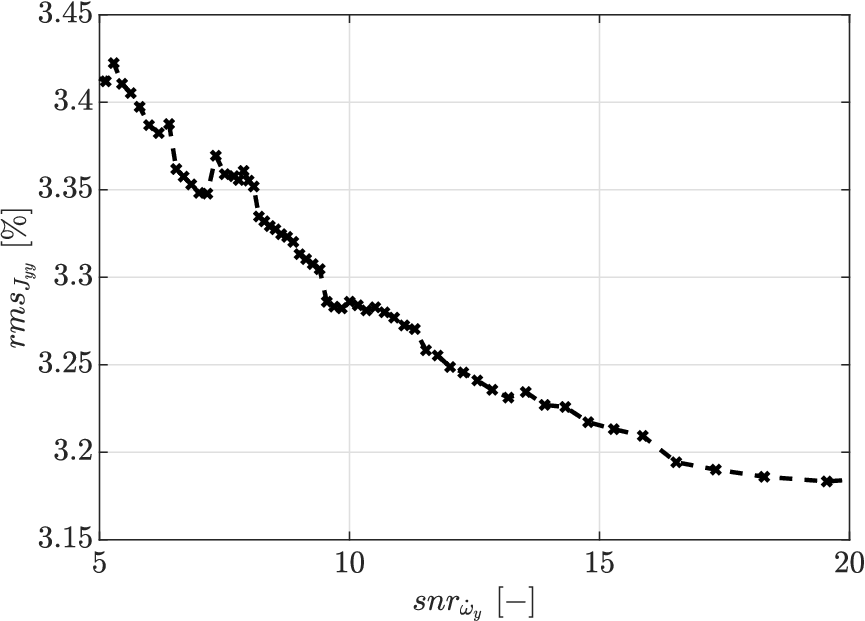}
		\caption{Pitch inertia estimation with varying noise levels. Oscillation in $rms_{\%}$ is intrinsically due to the complex nature of the Twin-in-the-Loop observer.}
		\label{fig:noise_sensitivity_Jyy}
	\end{minipage}
\end{figure}

\begin{figure}[th]
	\centering
	
	\subfloat[Sensitivity to $J_{yy}$ estimation error, based on the amount of noise in the road profile.\label{fig:pitch_inertia_sensitivity_to_road}]{\includegraphics[width=0.49\columnwidth]{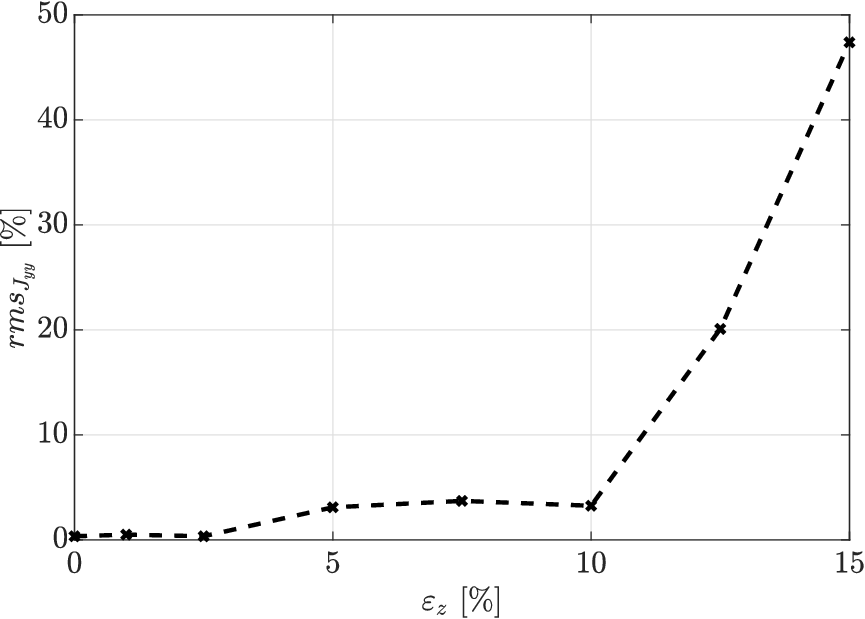}}
	\hfill
	\subfloat[Noisy versus real road profile, in case $\varepsilon_z=0.1$.\label{fig:road_noise}]{\includegraphics[width=0.49\columnwidth]{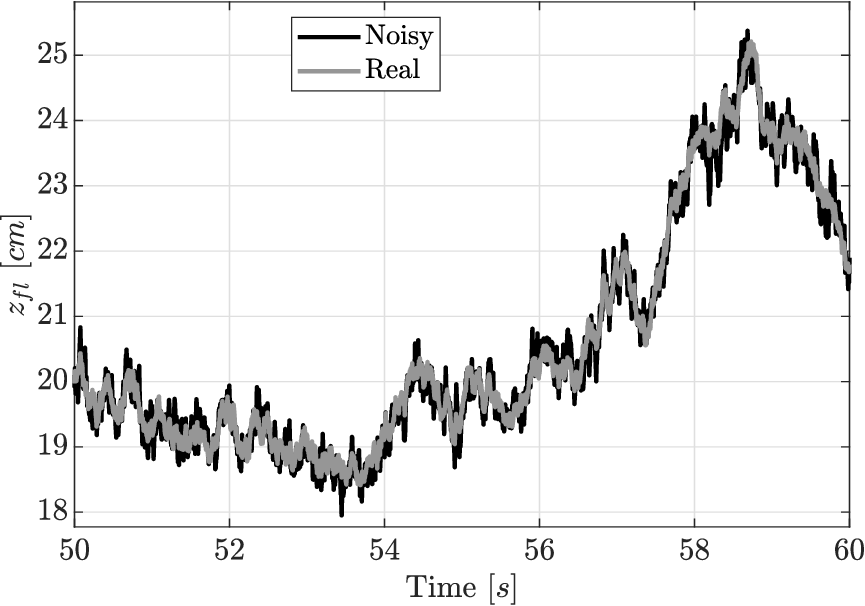}}
	
	\caption{Analysis on the effect of road profile noise onto the estimation of $J_{yy}$. (a) shows the sensitivity to $J_{yy}$ estimation error for different noise levels, while (b) compares noisy and real measurements for a specific noise level.}
	\label{fig:road_analysis}
\end{figure}

As done above, pitch inertia $J_{yy}$ can be estimated by suitably exciting the corresponding dynamics. The most straightforward way of exciting pitch dynamics is the excitation of the vehicle body with a non-zero road profile: hence, we assume that the vehicle is being driven at constant speed on a paved road. The road profile at front wheels is modeled as \citep{savaresi_2010}
\begin{equation}
	z_f = \int_{0}^{t} \eta_z,\ \eta_z \sim WN(0,\sigma_z^2),
\end{equation}
Where $\sigma_z=0.01\ m$, and $WN$ is the realization of a white noise. Consequently, the profile on the rear wheels is equal to $z_r(t)=z_f(t-v/wb)$, with $v$ the vehicle longitudinal speed, and $wb$ the wheelbase.

The real vehicle, whose load is unknown and has to be estimated, is fed with the generated road profile. The digital twin is then fed with the same road profile. In pratice, we are considering that the road profile at the current time is known; this information can be easily extracted \emph{e.g.} via Kalman Filtering \cite{qin_2017} or sensing devices \citep{theunissen_2020}. To achieve a realistic simulation, we assume the presence of noise in such information
\begin{equation}
	z_{f,k}^n=z_{f,k}\cdot u_{z,k},\ u_{z,k} \sim \mathcal{U}\left(1-\varepsilon_z,1+\varepsilon_z\right).
	\label{eq:noise_road}
\end{equation}
Equation \eqref{eq:noise_road} models a multiplicative uniformly distributed noise onto $z_f$.\\
The correction law to be applied is derived as usual
\begin{equation}
	\Delta \delta J_{yy,k_k}=K_{\dot{\omega}_{y}-\delta J_{yy}}(\dot{\omega}_{y,k}- \tilde{\dot{\omega}}_{y,k})\cdot \textrm{sgn}(\dot{\omega}_{y,k}),
	\label{eq:Jyy_correction}
\end{equation}
Where the $y$-axis angular acceleration $\dot{\omega}_y$ is employed as a measurement. In the following, only correction in Eq. \eqref{eq:Jyy_correction} is applied.

As also discussed for roll and yaw inertia, a preliminary estimation of the vehicle mass can yield benefits also to inertia estimation; the same goes for the center-of-mass position.
The inertia estimation - considering noisy measurements - are showed in Fig. \ref{fig:pitch_inertia_withMass_roadUncert_est}. The estimate converges to the correct value even by starting from different initial conditions. Also, note that the observer is robust with respect to imperfect knowledge of other inertia parameters $J_{xx}$ and $J_{zz}$, which are not being corrected.

Then, Table \ref{tab:pitch_table} quantifies the estimation performance for different simulation layouts: even in the worst scenario, \emph{i.e.} in case of noisy measurements and road profile information, and wrong knowledge of mass and center-of-mass location, the percentage $rms$ settles at $\approx 10\%$.

Finally, Fig. \ref{fig:pitch_inertia_sensitivity_to_road} shows the variation of $rms_{\%}$ for increasing road profile noise ($\varepsilon_z$): a significant performance drop arises for $\varepsilon_z>0.1$.

For a better understanding of the introduced noise level, Fig. \ref{fig:road_noise} depicts the real versus noisy road profiles. The $rms$ among the two signals is $\approx 0.2\ cm$, which is comparable to what observed in \cite{qin_2017}.
\begin{table}[th]
	\centering
	\caption{Pitch inertia estimation performance in different conditions of measurement noise, road profile noise, and with or without the mass estimation.}
	\label{tab:pitch_table}
	\begin{tabular}{|l|l|} 
		\hline
		\textbf{Test conditions}                                                                                                           &\boldmath{ $rms_{J_{yy}}\ [\%]$}  \\ 
		\hline
		\begin{tabular}[c]{@{}l@{}}Noiseless data ($snr=\infty$);\\Noiseless road profile ($\varepsilon_z=0$);\\Correct mass;\\Correct CM.\end{tabular} & $0.11$                \\ 
		\hline
		\begin{tabular}[c]{@{}l@{}}Noiseless data ($snr=\infty$);\\Noisy road profile ($\varepsilon_z=0.1$);\\Correct mass;\\Correct CM.\end{tabular}   & $3.19$                \\ 
		\hline
		\begin{tabular}[c]{@{}l@{}}Noisy data ($snr\approx10$);\\Noiseless road profile ($\varepsilon_z=0$);\\Correct mass;\\Correct CM\end{tabular}   & $0.33$                \\ 
		\hline
		\begin{tabular}[c]{@{}l@{}}Noisy data ($snr\approx 10$);\\Noisy road profile ($\varepsilon_z=0.1$);\\Correct mass;\\Correct CM\end{tabular}    & $3.24$                \\ 
		\hline
		\begin{tabular}[c]{@{}l@{}}Noisy data ($snr\approx 10$);\\Noisy road profile ($\varepsilon_z=0.1$);\\Wrong mass;\\Correct CM.\end{tabular}      & $2.38$                \\
		\hline
		\begin{tabular}[c]{@{}l@{}}Noisy data ($snr\approx 10$);\\Noisy road profile ($\varepsilon_z=0.1$);\\Correct mass;\\Wrong CM.\end{tabular}      & $5.51$                \\
		\hline
		\begin{tabular}[c]{@{}l@{}}Noisy data ($snr\approx 10$);\\Noisy road profile ($\varepsilon_z=0.1$);\\Wrong mass;\\Wrong CM.\end{tabular}      & $10.20$                \\
		\hline
	\end{tabular}
\end{table}
\subsection{Implementation details}
\label{Section:implementation_details}
\begin{figure}[th]
	\centering
	\includegraphics[width=0.7 \columnwidth]{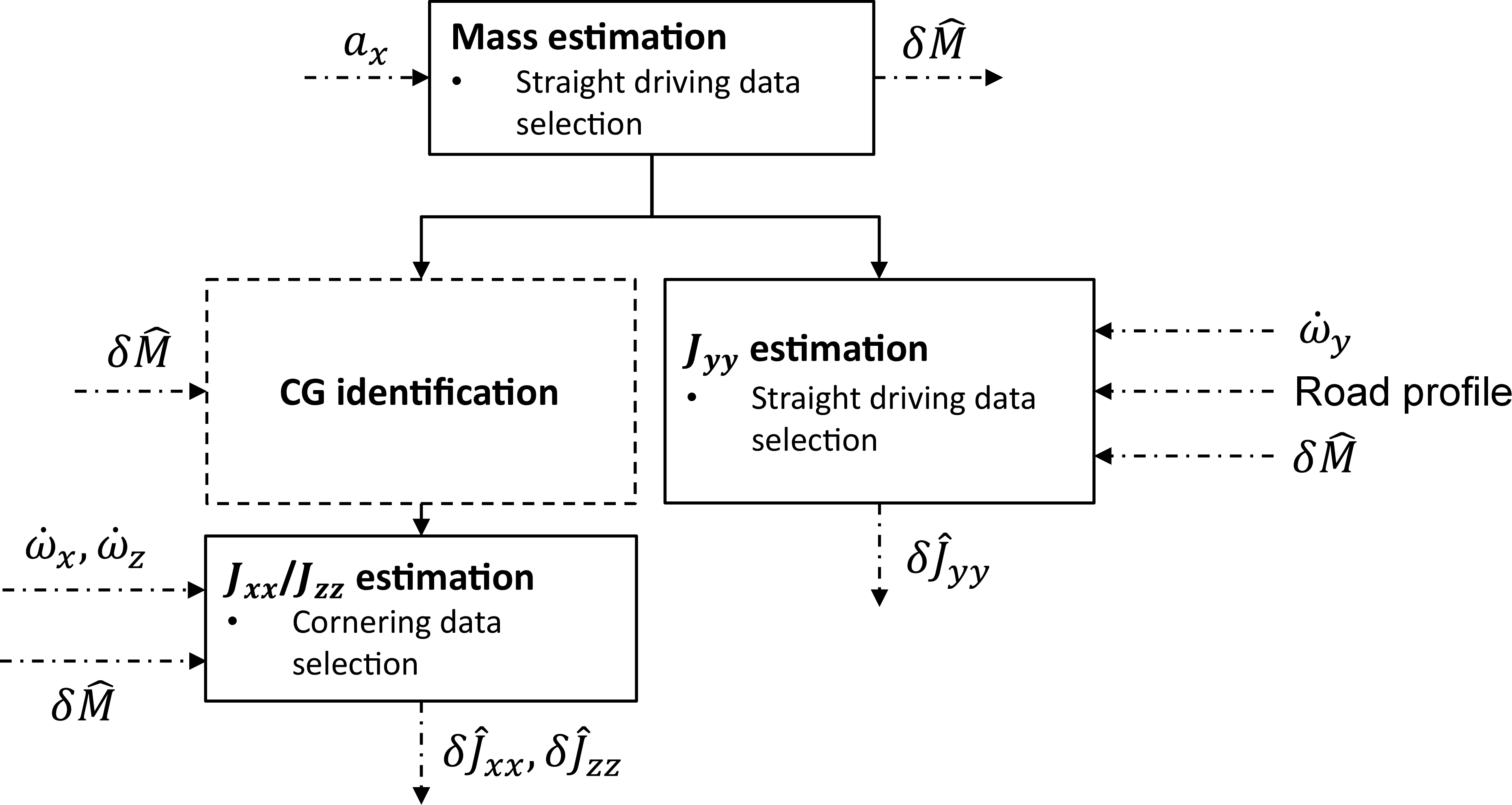}
	\caption{Schematic representation of the operations flow of a possible implementation of the TiL parameters estimator.}
	\label{fig:estimator_flow}
\end{figure}
In the Sections above, we showed how the TIL approach can properly estimate mass and inertia under different noise and uncertainty conditions; this allowed us to focus on each problem independently. In an actual implementation of the full parameter estimator, one could consider simple logic rules to prioritize estimation of one or more variables. A possibility is that of Fig. \ref{fig:estimator_flow}: the dashed arrows represent outputs/input of a given module of the estimator, while solid arrows describe the flow of operations.

The first stage consists in estimating the vehicle mass. We verified (see Table \ref{tab:mass_table}) as this part of the algorithm is extremely robust to knowledge of the center-of-mass and other inertial parameters; hence, this module is the first one to be executed. Estimation during straight driving can be enforced to avoid too highly dynamic conditions: to do so, a simple threshold on vehicle yaw-rate is sufficient. The transition to the next module can be based upon a time threshold: setting a reasonably high threshold ensures that the estimator has converged to the true value when transitioning.

Now, one could estimate the pitch inertia $J_{yy}$, which has been showed to be robust to center-of-mass knowledge (Table \ref{tab:pitch_table}); as discussed, this step requires knowledge of the road profile. Also in this case, enforcing straight driving can enhance robustness by avoiding parameter corrections in highly dynamic conditions.

On the other hand, in order to estimate other inertia parameters $J_{xx},\ J_{zz}$ we need to know the center-of-mass variation beforehand, so as to have reasonable estimation errors (within the $10\ \%$ bound). A suitable estimation procedure can thus be applied to solve for this problem; this enables estimation of roll and yaw inertia, for which a data selection to isolate cornering events is necessary - a threshold on yaw-rate or lateral acceleration solves the last point.

\section{Experimental validation}
\label{Section:case_study_experimental}
In the following, we test the parameter estimator onto real data. Data has been collected in a proving ground with an high-performance car, and provided by a partner car manufacturer whose name cannot be disclosed for confidentiality reasons.

The vehicle is provided with a set of sensors, yielding the following measurements
\begin{itemize}
	\item Center-of-mass 3D accelerations and angular rates (IMU);
	\item Four wheel angular rates (encoders).
\end{itemize}
For validation purposes, a double antenna GPS is used to measure longitudinal and lateral speed ($v_x$, $v_y$). Consequently, the vehicle sideslip $\beta$ can be computed as $\beta = atan(v_y/v_x)$.
We consider joint estimation of parameters and states: namely, we want to estimate vehicle sideslip and mass. Hence, the augmented state vector is
\begin{equation}
	x^{aug} = \begin{bmatrix} x^T & \delta M	\end{bmatrix}^T,
\end{equation}
where $x\in \mathbb{R}^{28\times 1}$ is the set of simulator states. \\
Measurements vector $y\in \mathbb{R}^{10\times 1}$ reads
\begin{equation}
		y=\begin{bmatrix}a_x & a_y & a_z& \omega_x & \omega_y & \omega_z & \omega_{fl} & \omega_{fr} & \omega_{rl} & \omega_{rr}	\end{bmatrix}^T
\end{equation}
The most relevant vehicle nominal parameters are collected in Table \ref{tab:experimental_vehicle_parameters}.
\begin{table*}[]
	\centering
	\begin{tabular}{|lllllll|}
		\hline
		\multicolumn{7}{|c|}{\textbf{Parameter}}                                                                                                                                                                                           \\ \hline
		\multicolumn{1}{|l|}{$M_{tot}$} & \multicolumn{1}{l|}{$J_{xx}$} & \multicolumn{1}{l|}{$J_{yy}$} & \multicolumn{1}{l|}{$J_{zz}$} & \multicolumn{1}{l|}{$d_x$ {[}$cm${]}} & \multicolumn{1}{l|}{$d_y$ {[}$cm${]}} & $d_z$ {[}$cm${]} \\ \hline
		\multicolumn{1}{|l|}{$1391.29$}    & \multicolumn{1}{l|}{$328.281$}    & \multicolumn{1}{l|}{$1698.21$}   & \multicolumn{1}{l|}{$1864.95$}   & \multicolumn{1}{l|}{$157.14$}             & \multicolumn{1}{l|}{$0$}              & $46.22$             \\ \hline
	\end{tabular}
	\caption{Test vehicle nominal parameters, as modeled on the digital twin.}
	\label{tab:experimental_vehicle_parameters}
\end{table*}
\subsection{Tuning the TiL observer}
\begin{algorithm2e}[th]
	\caption{\rule[0ex]{0pt}{2ex}BO for TiL calibration: pseudo-code\rule[-0.6ex]{0pt}{1ex}}
	\label{alg:BO}
	\SetNlSty{text}{}{:}
	\SetAlgoNlRelativeSize{-1}
	\DontPrintSemicolon
	Select an experiment, consisting of measured states, input, measurements $\left(x_k,u_k,y_k\right)$\;
	Select $\mathcal{N}$ number of total iterations, $n_{seed}$ number of initial points ($n_{seed}<\mathcal{N}$)\;
	Evaluate the objective function in $n_{seed}$ random initial points\;
	$i \leftarrow n_{seed}$\;
	\While{$i < \mathcal{N}$}
	{
		Update surrogate function $\hat{f}_i(\theta)$\label{step:surrogate_function}\;
		Compute acquisition function $a_i(\theta)$\label{step:acquisition_function}\;
		Next point to evaluate is $\theta_{i+1}=\arg\min_{{\theta}} a_i(\theta)$\label{step:min_acquisition_function}\;
		$i\leftarrow i+1$\;
		Evaluate $y_i=f(\theta_i)$\;
	}
	Update surrogate function $\hat{f}_N(\theta)$\;
	\KwRet{\upshape 1) best evaluated point $\overline{\theta} = \arg\min_i f(\theta_i)$}\;
	\vspace{-1pt} \nonl 2) best predicted feasible point $\hat{\overline{\theta}}=\arg\min_\theta \hat{f}_{\mathcal{N}}(\theta)$
\end{algorithm2e}
Applying the correction in Eq. \eqref{eq:correction_nonlinear} with set of states and measurements defined above would yield to state-output mapping matrix $K\in \mathbb{R}^{28\times 10}$. With regards to the non-linear correction $\mathcal{K}$, to be applied for the estimation, the expression in Eq. \eqref{eq:mass_correction} is used.
All considered literature solutions rely on Kalman Filters to compute the correction gain to be applied: unfortunately, given that the digital twin is a black-box, we cannot access its equations, and thus we cannot employ standard techniques - \emph{e.g.} the system linearization, as in Extended Kalman Filter. A data-driven approach is instead here proposed, following what showed in \citep{riva_2022_SIL}. An optimization procedure is performed based on experimental data: a set of states, driver inputs and measurements is necessary $\langle x_k,u_k,y_k\rangle$. Such procedure calibrates the parameters off-line by minimizing a certain cost function - depending on the estimator objective.

Given that the cost metric to be optimized cannot be written in closed-form - being the simulator equation unknown - we need to rely on a black-box optimizer.
These methods suffer of huge scalability problems, and the number of optimization variables shall be reduced as much as possible; this can be enforced by removing variables from the measurements and states vector, reducing the matrix order. Specifically, we consider the following subset of $y$
\begin{equation}
	y_{ss} = \begin{bmatrix}
		a_x & a_y  & \omega_y & \omega_z & \omega_{fl} & \omega_{fr} & \omega_{rl} & \omega_{rr}
	\end{bmatrix}^T.
\end{equation}
$a_z$ and $\omega_x$ have been removed from the set, as we are interested in estimating planar dynamics; on the other hand, one can verify that including the pitch rate $\omega_y$ among the measurements can enhance the estimation performance, as the variable is highly related to the road profile - and thus to suspension forces.

The same considerations can be applied to the state vector, thus obtaining a reduced version
\begin{equation}
		x_{ss}=\begin{bmatrix}v_x & v_y  & \omega_y & \omega_z & \omega_{fl} & \omega_{fr} & \omega_{rl} & \omega_{rr} & \delta M
		\end{bmatrix}^T.
\end{equation}
Where the planar velocities allow to algebraically estimate $\beta$; wheel, pitch and yaw angular rates $\omega_{fl,fr,rl,rr},\ \omega_y,\ \omega_z$ are corrected with the corresponding measurements in order to guarantee simulator stability.
At this point, mapping matrix has dimensions $9\times 8$, meaning that $72$ parameters shall be calibrated. $72$ parameters are still computationally intractable, hence, we promote physics-inspired sparsity in the matrix \citep{delcaro_2023}. In details, we consider the following corrections
\begin{itemize}
	\item $K_{\omega - \omega}$. Correction onto the wheel angular rates via the corresponding measurements;
	\item $K_{a_x - v_x}$ and $K_{a_y \rightarrow v_y}$. Correction onto the vehicle longitudinal/lateral speed via the corresponding accelerations;
	\item $K_{\omega_z - \omega_z}$. Correction onto the yaw rate via the corresponding measurement;
	\item $K_{\omega_y - \omega_y}$. Correction onto the pitch rate via the corresponding measurement.
\end{itemize}
Hence, the set of parameters to be tuned reads
\begin{equation}
		\theta = \begin{bmatrix}
			K_{\omega - \omega} & K_{a_x - v_x}  & K_{a_y - v_y} & K_{\omega_y - \omega_y} & K_{\omega_z - \omega_z} & K_{a_x - \delta M}
		\end{bmatrix}
\end{equation}
Where the term $K_{a_x - \delta M}$ is the correction on the vehicle mass; we apply it through the switching law
\begin{equation}
	\begin{split}
		\Delta \delta M_k&=K_{a_{x}-\delta M}\cdot  \epsilon\left(a_{x,k},\omega_{z,k}\right)\cdot (a_{x,k}- \tilde{a}_{x,k}),\\
		\epsilon&=	\begin{array}{c}
			\begin{cases}
				0 &,\ \left|\omega_{z,k}\right|>\bar{\omega}_z,\\
				1 &,\ \textrm{sgn}\left(a_{x,k}\right)\geq0 \wedge \left|\omega_{z,k}\right|<\bar{\omega}_z, \\
				-1 &,\ \textrm{sgn}\left(a_{x,k}\right)<0 \wedge \left|\omega_{z,k}\right|<\bar{\omega}_z.
		\end{cases}\end{array}
	\end{split}
	\label{eq:mass_correction_implemented}
\end{equation}
Equation \eqref{eq:mass_correction_implemented} is very similar to Eq. \eqref{eq:mass_correction}: however, we introduce a scheduling law based on the yaw-rate $\omega_z$, with $\bar{\omega}_z=3\ deg/s$ - as also discussed in Section \ref{Section:implementation_details} - in order to perform the estimation during straight driving.
Now, the optimal observer calibration $\theta^*$ can be found by solving the following optimization problem
\begin{equation}
	\begin{split}
		\min_{{\theta}}\ &J\left(\theta\right)\\
		\textrm{subject to} &\ \theta \subseteq	\Theta.
	\end{split}
	\label{eq:optimization_problem}
\end{equation}
The cost function has to be selected based on the estimation targets. In this case, we have
\begin{equation}
		J(\theta) = \sqrt{\dfrac{1}{N_{s}}\sum_{i=1}^{N_{s}} k_\beta \left(\beta_k-\hat{\beta}_k\right)^2}+\sqrt{\dfrac{1}{10f_s}\sum_{i=N_s-10f_s}^{N_{s}} \left(\delta M_k -\hat{\delta M}_k\right)^2}.
\end{equation}
Whereas $k_\beta=100$ accounts for numeric differences among the two cost function terms, and $N_s$ is the number of samples in the experiment. Note that the second term of the cost function weights the mass in the last $10$ seconds of experiment. This is because we don't need and don't expect the mass estimation to be as fast as the sideslip one, hence we don't give great relevance to the first part of experiment.

We select Bayesian Optimization (BO) to solve the problem in Eq. \eqref{eq:optimization_problem}: BO is a model-free optimizer employed when the cost function to be minimized cannot be written down in form of equation, but we are able to evaluate for some values of $\theta$. A Gaussian Process proxy of the cost function is then found by regression on said points, and used to estimate the global optimum of the problem in Eq. \eqref{eq:optimization_problem}. The BO procedure is schematized in Algorithm \ref{alg:BO}, and more information can be found in one of the many papers about this topic, \emph{e.g.} \citep{gelbart_2014}.
\subsection{Benchmark estimator}
\label{Section:benchmark}
\begin{figure}[h]
	\centering
	\includegraphics[width=0.7 \columnwidth]{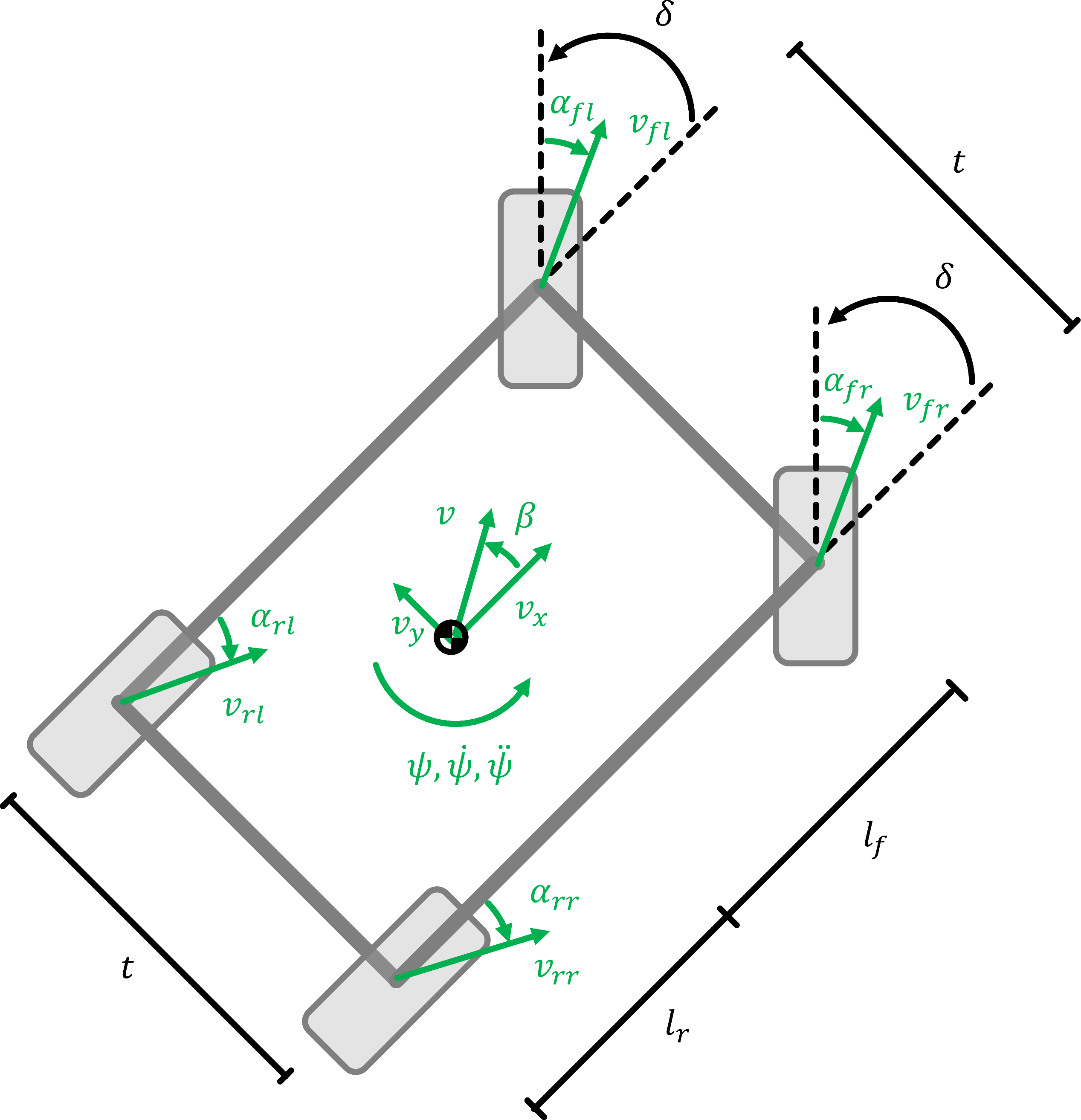}
	\caption{Double-track vehicle model scheme.}
	\label{fig:benchmark_scheme}
\end{figure}
In order to fairly compare the TiL estimator with a benchmark, we build an observer with a double-track planar model as a system replica (see Fig. \ref{fig:benchmark_scheme}); the same correction architecture as for the TiL, based on constant gains, is retained. This model has been frequently used in the state and parameter estimation literature (\citep{hong_2014,zhu_2019}). Furthermore, we use double-track model for consistency to our previous research on TiL topic \citep{riva_2022_SIL}. \\ The planar vehicle dynamics are written - in discrete time - as
\begin{equation}
		\begin{split}
			v_{x,k+1}&=v_{x,k}+T_s\left(\dfrac{F_x^T}{M_{tot}+\delta M_k}+v_{y,k} \omega_{z,k}\right), \\
			v_{y,k+1}&=v_{y,k}+T_s\left(\dfrac{F_y^T}{M_{tot}+\delta M_k}-v_{x,k} \omega_{z,k}\right), \\
			\omega_{z,k+1}&=\omega_{z,k}+T_s\left(\dfrac{M_z^T}{J_{zz}}\right), \\
			\delta M_{k+1}&=\delta M_k.
			\label{eq:double_track_eqs}
		\end{split}
\end{equation}
Where $T_s$ is the sampling time. The equations are enhanced with the extended state for the mass estimation, $\delta M$.
The total forces $F_x^T,\ F_y^T$ and moment $M_z^T$ are respectively written as a function of the tire-road forces 
\begin{equation}
		\begin{split}
			F_x^T &= \left(F_{x,fl}+F_{x,fr}\right)\cos(\delta)-(F_{y,fl}+F_{y,fr})\sin(\delta)+\\ 
			&F_{x,rl}+F_{x,rr}, \\
			F_y^T &= \left(F_{x,fl}+F_{x,fr}\right)\sin(\delta)+(F_{y,fl}+F_{y,fr})\cos(\delta)+\\
			&F_{y,rl}+F_{y,rr}, \\
			M_z^T &= l_f \left(F_{y,fl}+F_{y,fr}\right)\cos(\delta)+\dfrac{t}{2}(F_{y,fl}-F_{y,fr})\sin(\delta)-\\
			&\dfrac{t}{2}(F_{x,fl}-F_{x,fr})\cos(\delta)+l_f\left(F_{x,fl}+F_{x,fr}\right)\sin(\delta)-\\
			&l_r\left(F_{y,rl}+F_{y,rr}\right)-\dfrac{t}{2}\left(F_{x,rl}-F_{x,rr}\right).
		\end{split}
\end{equation}
Where $l_f$,\ $l_r$,\ $t$ are parameters influenced by the center-of-mass position, which could thus change over time. However, given that the considered case study is a 2-seats car, the expected load variations are not of significant influence to the CM position.

The wheel forces $F_{x,ij},\ F_{y,ij}$ are modeled through a simplified Pacejka model \cite{pacejka_2005}
\begin{equation}
	\begin{split}
		F_x^{ij}&=F_{z,ij}D_x \sin\left(C_x \arctan\left(B_x \lambda_{ij}-E_x\left(B_x \lambda_{ij}-\arctan\left(B_x \lambda_{ij}\right)\right)\right)\right), \\
		F_y^{ij}&=F_{z,ij}D_y \sin\left(C_y \arctan\left(B_y \alpha_{ij}-E_y\left(B_y \alpha_{ij}-\arctan\left(B_y \alpha_{ij}\right)\right)\right)\right).
	\end{split}
	\label{eq:pacejka}
\end{equation}
Where $F_z^{ij}$ is the normal force at each wheel; the same can be easily estimated by using center-of-mass measured accelerations \citep{viehweger_2021}. The wheel longitudinal and lateral slips $\lambda_{ij},\ \alpha_{ij}$ are computed as
\begin{equation}
	\begin{split}
		\lambda_{ij}&=\dfrac{R_{ij}\omega_{ij}-v_{x,ij}}{\max \left(R_{ij}\omega_{ij},v_{x,ij}\right)}, \\
		\alpha_{ij}&=\arctan\left(v_y^{ij}/v_x^{ij}\right).
	\end{split}
	\label{eq:wheel_slips}
\end{equation}
Where $v_{x,ij},\ v_{y,ij}$ are obtained from simple kinematic considerations.
In principle, one could include the wheel dynamics as in \citep{riva_2022_SIL}, in order to filter the wheel angular rate states. However, since the estimation target is different, and we assume the angular rate measurements to be reliable, we directly consider measured $\omega_{ij}$ in Eq. \eqref{eq:wheel_slips}. This also avoids us to model the braking and traction torques effect onto the wheels, a potential further source of error. 
Overall, Eq. \eqref{eq:double_track_eqs}, Eq. \eqref{eq:pacejka}, Eq. \eqref{eq:wheel_slips} can be combined in order to obtain the double-track dynamics
\begin{equation}
		\begin{split}
			x_{bench,k+1}&=x_{bench,k}+T_s\cdot f_{bench}\left(x_{bench,k},u_{bench,k}\right), \\
			y_{bench,k}&=\begin{bmatrix}
				a_{x,k} \\ a_{y,k} \\ \omega_{z,k}
			\end{bmatrix}=\begin{bmatrix}
				\dfrac{F_x^T}{M_{tot}+\delta M_k}+v_{y,k} \omega_{z,k} \\ \dfrac{F_y^T}{M_{tot}+\delta M_k}-v_{x,k} \omega_{z,k} \\ \omega_{z,k}
			\end{bmatrix}.
		\end{split}
\end{equation}
Whereas $x_{bench}=\begin{bmatrix}
	v_x & v_y & \dot{\psi} & \delta M
\end{bmatrix}^T\in \mathbb{R}^4$, and $u_{bench,k}=\begin{bmatrix}
	\delta & \omega_{fl} & \omega_{fr} & \omega_{rl} & \omega_{rr}
\end{bmatrix}\in \mathbb{R}^5$. $f_{bench}$ is a suitable non-linear function. \\
We apply on the benchmark estimator the same correction law proposed in Section \ref{Section:Problem_statement}. The correction gains are stored in vector $\theta_{bench}$
\begin{equation}
		\theta_{bench}=\begin{bmatrix}
			K_{a_x-v_x} & K_{\omega_z - \omega_z} & K_{a_x - \delta M} & K_{a_y - v_y}
		\end{bmatrix}.
\end{equation}
Bayesian Optimization is employed to tune $\theta_{bench}$, as for the TiL observer.

\begin{remark}
	Let us remark that the double-track model - especially the Pacejka parameters - need to be identified based on experimental data: this increases the number of tunable parameters and the calibration complexity of the benchmark. On the other hand, a digital twin of the vehicle is usually already well calibrated, as the same is used for simulation purposes by the car manufacturer. Furthermore, the double-track model is not valid anymore if we want to estimate roll or pitch inertia: we would need to design another ad-hoc model, or to enhance the double-track with non-planar dynamics.
\end{remark}
\subsection{Final results}
\begin{figure}[th]
	\centering
	\begin{minipage}[b]{0.49\columnwidth}
		\centering
		\includegraphics[width=\linewidth]{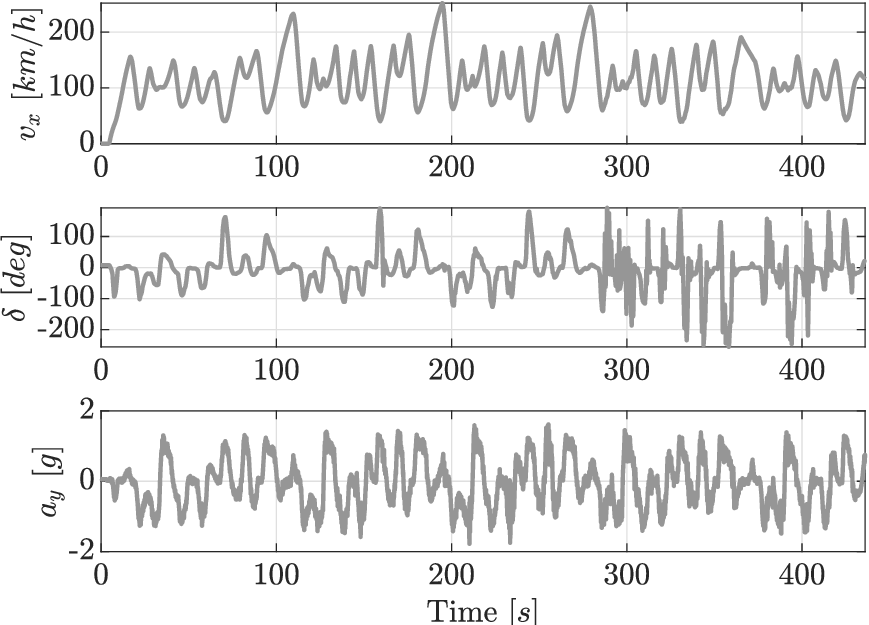}
		\caption{Speed, yaw-rate and lateral acceleration profiles in a series of circuit laps (optimization experiment).}
		\label{fig:optimization_experiment}
	\end{minipage}
	\hfill
	\begin{minipage}[b]{0.49\columnwidth}
		\centering
		\includegraphics[width=\linewidth]{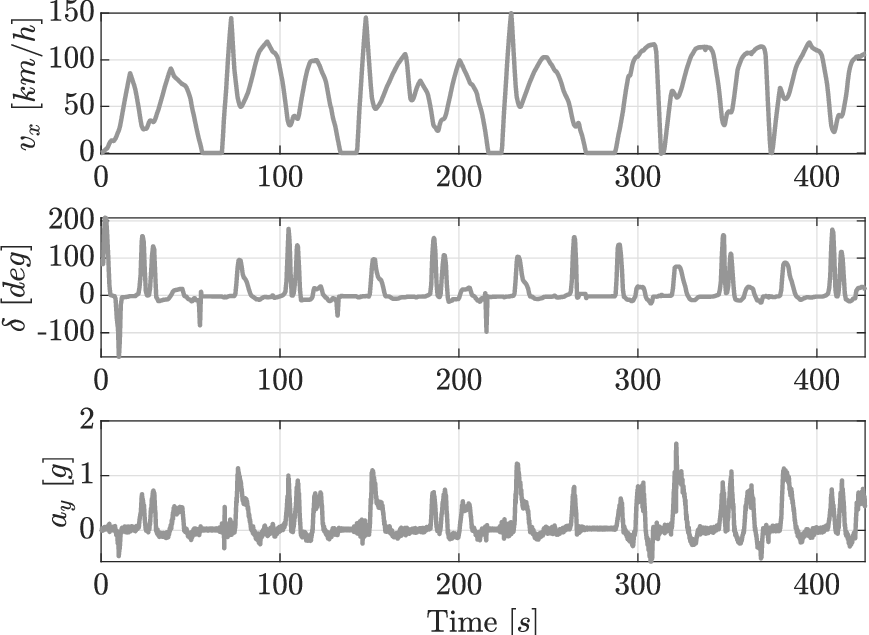}
		\caption{Speed, yaw-rate and lateral acceleration profiles in a series of double-lane-change maneuvers (validation experiment).}
		\label{fig:validation_experiment}
	\end{minipage}
\end{figure}

\begin{figure}[th]
	\centering
	\includegraphics[width=1\columnwidth]{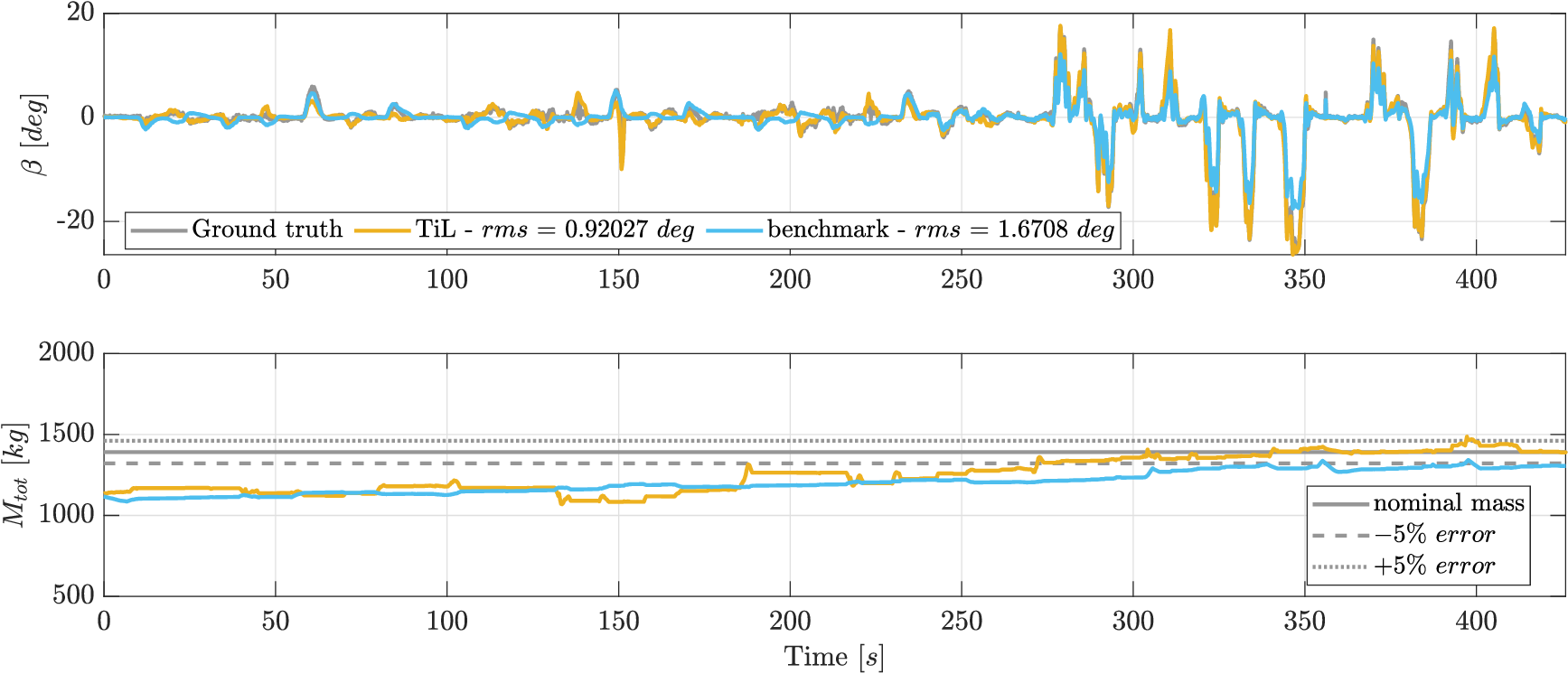}
	\caption{Joint mass and sideslip estimation in a series of circuit laps (optimization experiment). The TIL estimator is compared with a benchmark estimator.}
	\label{fig:mass_sideslip_est}
\end{figure}

\begin{figure}[th]
	\centering
	\includegraphics[width=0.8\columnwidth]{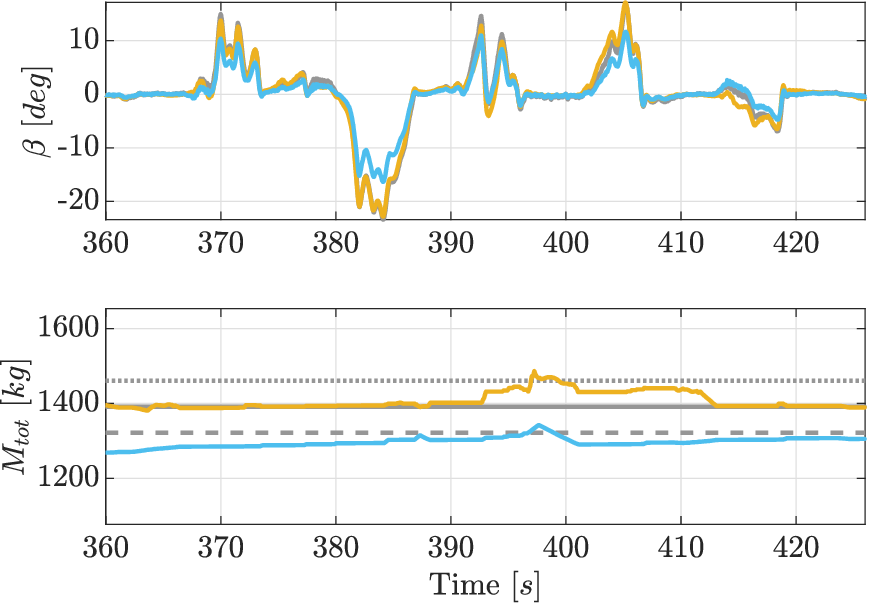}
	\caption{Joint mass and sideslip estimation in a series of circuit laps (optimization experiment). The TIL estimator is compared with a benchmark estimator - highlighted portion.}
	\label{fig:mass_sideslip_est_zoom}
\end{figure}

\begin{figure*}[th]
	\centering
	\includegraphics[width=1 \columnwidth]{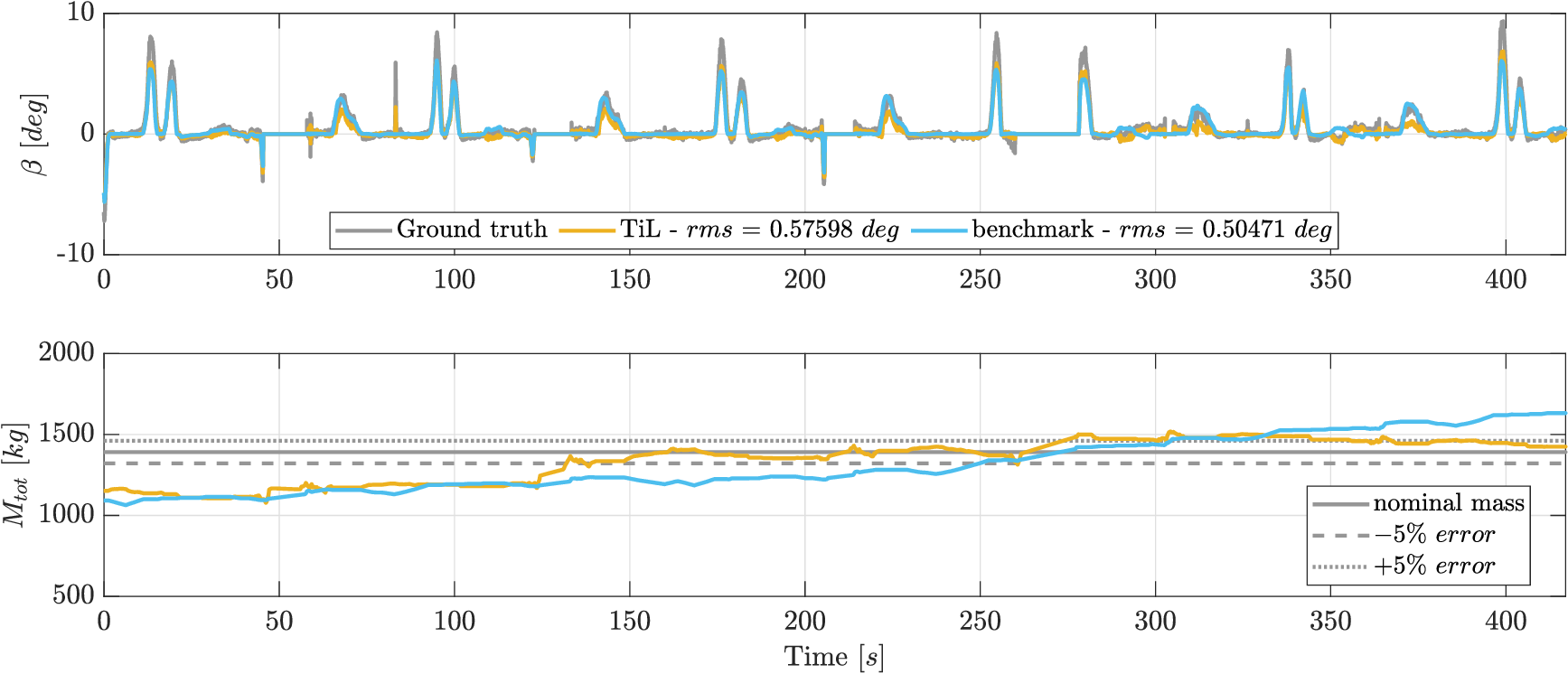}
	\caption{Joint mass and sideslip estimation in a series of circuit laps (validation experiment). The TIL estimator is compared with a benchmark estimator.}
	\label{fig:mass_sideslip_est_validation}
\end{figure*}
\begin{table*}[th]
	\begin{tabular}{|l|c|c|c|c|}
		\hline
		\multicolumn{1}{|c|}{\textbf{Parameter}} & \textbf{Lower bound} & \multicolumn{1}{l|}{\textbf{Upper bound}} & \multicolumn{1}{l|}{\textbf{Optimized - TiL}} & \multicolumn{1}{l|}{\textbf{Optimized - bench}} \\ \hline
		$k_{\omega-\omega}$                      & $0$                  & $1.5$                                     & $0.259$                                         & /                                               \\ \hline
		$k_{a_x-v_x}$                            & $0.01$               & $0.01$                                    & $3.287\cdot 10^{-3}$                                         & $9.282\cdot 10^{-2}$                                           \\ \hline
		$k_{a_y-v_y}$                            & $0.01$               & $0.01$                                    & $7.430\cdot 10^{-2}$                                         & $5.271\cdot 10^{-2}$                                           \\ \hline
		$k_{\omega_y-\omega_y}$                  & $0.2$                & $1.5$                                     & $0.913$                                       & /                                               \\ \hline
		$k_{\omega_z-\omega_z}$                  & $0.2$                & $1.5$                                     & $1.046$                                       &                     $0.314$                            \\ \hline
		$k_{a_x-\delta M}$                       & $0$                  & $1000$                                    & $248.17$                                         &                $2.06$                       \\ \hline
	\end{tabular}
	\caption{Upper, lower bounds and optimal parameters found for the TiL estimator via BO.}
	\label{tab:optimal_paarameters_TIL}
\end{table*}
The following results are obtained by testing the TiL estimator and the benchmark defined in Section \ref{Section:benchmark}. The experiments have been performed on the vehicle having a nominal mass, hence, in order to test the algorithms, we modify the initialization of the estimator internal models, so to have a wrong mass value $\delta_{M,0}=-350\ kg$.

Figure \ref{fig:optimization_experiment} depicts vehicle speed, steering wheel angle and lateral acceleration in the experiment used for solving optimization problem in Eq. \eqref{eq:optimization_problem} - the test consists in a series of circuit laps. As the reader can note, the experiment is extremely challenging, in that accelerations up to $1.5\ g$ and speed up to $230\ km/h$ are reached.

For this test, we apply the BO procedure solving the problem of Eq. \eqref{eq:optimization_problem}: the optimal parameters in Table \ref{tab:optimal_paarameters_TIL} are obtained. The variable bounds can be found via trial-and-error and prior physical knowledge, as discussed in \citep{riva_2022_SIL}
The estimation results are displayed in Fig. \ref{fig:mass_sideslip_est}. An highlight on the last part of the test is showed in Fig. \ref{fig:mass_sideslip_est_zoom}. As one can notice, TiL is able to outperform the double-track based observer both on mass and sideslip estimation. The mass estimate convergence is indeed slower than in the simulation test of Section \ref{section:vehicle_mass_estimation}: this is however expected, as the real world experiment is significantly more challenging. 

Figure \ref{fig:validation_experiment} shows speed, yaw-rate and lateral acceleration for another experiment, used for validating the estimators - this test consists in a series of double lane change and braking maneuvers, and is thus different in nature from the optimization one. The estimation results are given in Fig. \ref{fig:mass_sideslip_est_validation}. Also in this case, both filters are able to properly estimate the sideslip angle - with a comparable performance. On the other hand, the double-track model is not able to correctly estimate the mass, and seems to be diverging over time. 
\section{Conclusions}
\label{Section:Conclusions}
In this manuscript, we show that the mass and inertia of a vehicle can be estimated  - jointly with classical vehicle dynamics states - via the recently introduced Twin-in-the-Loop filtering approach.
The method is validated extensively in simulation, considering independent estimation of each parameter of interest, and performing sensitivity analyses to noise and uncertainty.
Then, the TiL estimator is tested against experimental data collected on a high-performance car, showing high performance in joint estimation of mass and sideslip. A comparison with another observer based on state-of-the-art double-track vehicle modelling provides further insights into the potential of the TiL architecture.

Future work will be dedicated to the study of the real-time estimation of the tire-road friction coefficient.
\section*{Conflicts of Interest}
The authors declare no conflict of interest.

\bibliographystyle{elsarticle-harv}

\bibliography{ref}

\begin{thebibliography}{27}
\expandafter\ifx\csname natexlab\endcsname\relax\def\natexlab#1{#1}\fi
\providecommand{\url}[1]{\texttt{#1}}
\providecommand{\href}[2]{#2}
\providecommand{\path}[1]{#1}
\providecommand{\DOIprefix}{doi:}
\providecommand{\ArXivprefix}{arXiv:}
\providecommand{\URLprefix}{URL: }
\providecommand{\Pubmedprefix}{pmid:}
\providecommand{\doi}[1]{\href{http://dx.doi.org/#1}{\path{#1}}}
\providecommand{\Pubmed}[1]{\href{pmid:#1}{\path{#1}}}
\providecommand{\bibinfo}[2]{#2}
\ifx\xfnm\relax \def\xfnm[#1]{\unskip,\space#1}\fi
\bibitem[{Delcaro et~al.(2023)Delcaro, Dettù, Formentin and
  Savaresi}]{delcaro_2023}
\bibinfo{author}{Delcaro, G.}, \bibinfo{author}{Dettù, F.},
  \bibinfo{author}{Formentin, S.}, \bibinfo{author}{Savaresi, S.},
  \bibinfo{year}{2023}.
\newblock \bibinfo{title}{Dealing with the curse of dimensionality in
  twin-in-the-loop observer design}, in: \bibinfo{booktitle}{2023 IFAC World
  Congress}, \bibinfo{organization}{IFAC}.
\bibitem[{Dettù et~al.(2022)Dettù, Formentin and Savaresi}]{dettu_2022_SILC}
\bibinfo{author}{Dettù, F.}, \bibinfo{author}{Formentin, S.},
  \bibinfo{author}{Savaresi, S.M.}, \bibinfo{year}{2022}.
\newblock \bibinfo{title}{The twin-in-the-loop approach for vehicle dynamics
  control}.
\newblock \bibinfo{journal}{arXiv:2209.02263} .
\bibitem[{Gelbart et~al.(2014)Gelbart, Snoek and Adams}]{gelbart_2014}
\bibinfo{author}{Gelbart, M.A.}, \bibinfo{author}{Snoek, J.},
  \bibinfo{author}{Adams, R.P.}, \bibinfo{year}{2014}.
\newblock \bibinfo{title}{Bayesian optimization with unknown constraints}.
\newblock \bibinfo{journal}{arXiv preprint arXiv:1403.5607} .
\bibitem[{Gevers(2005)}]{gevers2005identification}
\bibinfo{author}{Gevers, M.}, \bibinfo{year}{2005}.
\newblock \bibinfo{title}{Identification for control: From the early
  achievements to the revival of experiment design}.
\newblock \bibinfo{journal}{European journal of control} \bibinfo{volume}{11},
  \bibinfo{pages}{335--352}.
\bibitem[{Gobbi et~al.(2011)Gobbi, Mastinu and Previati}]{gobbi_2011}
\bibinfo{author}{Gobbi, M.}, \bibinfo{author}{Mastinu, G.},
  \bibinfo{author}{Previati, G.}, \bibinfo{year}{2011}.
\newblock \bibinfo{title}{A method for measuring the inertia properties of
  rigid bodies}.
\newblock \bibinfo{journal}{Mechanical Systems and Signal Processing}
  \bibinfo{volume}{25}, \bibinfo{pages}{305--318}.
\bibitem[{Gong et~al.(2020)Gong, Suh and Lin}]{gong_2020}
\bibinfo{author}{Gong, X.}, \bibinfo{author}{Suh, J.}, \bibinfo{author}{Lin,
  C.}, \bibinfo{year}{2020}.
\newblock \bibinfo{title}{A novel method for identifying inertial parameters of
  electric vehicles based on the dual h infinity filter}.
\newblock \bibinfo{journal}{Vehicle System Dynamics} \bibinfo{volume}{58},
  \bibinfo{pages}{28--48}.
\bibitem[{Hong et~al.(2014)Hong, Lee, Borrelli and Hedrick}]{hong_2014}
\bibinfo{author}{Hong, S.}, \bibinfo{author}{Lee, C.},
  \bibinfo{author}{Borrelli, F.}, \bibinfo{author}{Hedrick, J.K.},
  \bibinfo{year}{2014}.
\newblock \bibinfo{title}{A novel approach for vehicle inertial parameter
  identification using a dual kalman filter}.
\newblock \bibinfo{journal}{IEEE Transactions on Intelligent Transportation
  Systems} \bibinfo{volume}{16}, \bibinfo{pages}{151--161}.
\bibitem[{Huang and Wang(2014)}]{huang_2014}
\bibinfo{author}{Huang, X.}, \bibinfo{author}{Wang, J.}, \bibinfo{year}{2014}.
\newblock \bibinfo{title}{Real-time estimation of center of gravity position
  for lightweight vehicles using combined akf–ekf method}.
\newblock \bibinfo{journal}{IEEE Transactions on Vehicular Technology}
  \bibinfo{volume}{63}, \bibinfo{pages}{4221--4231}.
\newblock \DOIprefix\doi{10.1109/TVT.2014.2312195}.
\bibitem[{Li et~al.(2016)Li, Lu, Wang and Chen}]{li_2016}
\bibinfo{author}{Li, L.}, \bibinfo{author}{Lu, Y.}, \bibinfo{author}{Wang, R.},
  \bibinfo{author}{Chen, J.}, \bibinfo{year}{2016}.
\newblock \bibinfo{title}{A three-dimensional dynamics control framework of
  vehicle lateral stability and rollover prevention via active braking with
  mpc}.
\newblock \bibinfo{journal}{IEEE Transactions on Industrial Electronics}
  \bibinfo{volume}{64}, \bibinfo{pages}{3389--3401}.
\bibitem[{Pacejka(2005)}]{pacejka_2005}
\bibinfo{author}{Pacejka, H.}, \bibinfo{year}{2005}.
\newblock \bibinfo{title}{Tire and vehicle dynamics}.
\newblock \bibinfo{publisher}{Elsevier}.
\bibitem[{Qin et~al.(2017)Qin, Langari, Wang, Xiang and Dong}]{qin_2017}
\bibinfo{author}{Qin, Y.}, \bibinfo{author}{Langari, R.},
  \bibinfo{author}{Wang, Z.}, \bibinfo{author}{Xiang, C.},
  \bibinfo{author}{Dong, M.}, \bibinfo{year}{2017}.
\newblock \bibinfo{title}{Road profile estimation for semi-active suspension
  using an adaptive kalman filter and an adaptive super-twisting observer}, in:
  \bibinfo{booktitle}{2017 American Control Conference (ACC)},
  \bibinfo{organization}{IEEE}. pp. \bibinfo{pages}{973--978}.
\bibitem[{Reina et~al.(2017)Reina, Paiano and Blanco-Claraco}]{reina_2017}
\bibinfo{author}{Reina, G.}, \bibinfo{author}{Paiano, M.},
  \bibinfo{author}{Blanco-Claraco, J.L.}, \bibinfo{year}{2017}.
\newblock \bibinfo{title}{Vehicle parameter estimation using a model-based
  estimator}.
\newblock \bibinfo{journal}{Mechanical Systems and Signal Processing}
  \bibinfo{volume}{87}, \bibinfo{pages}{227--241}.
\bibitem[{Riva et~al.(2022)Riva, Formentin, Corno and Savaresi}]{riva_2022_SIL}
\bibinfo{author}{Riva, G.}, \bibinfo{author}{Formentin, S.},
  \bibinfo{author}{Corno, M.}, \bibinfo{author}{Savaresi, S.M.},
  \bibinfo{year}{2022}.
\newblock \bibinfo{title}{Twin-in-the-loop state estimation for vehicle
  dynamics control: theory and experiments}.
\newblock \bibinfo{journal}{arXiv:2204.06259} .
\bibitem[{Rodr{\'\i}guez et~al.(2021)Rodr{\'\i}guez, Sanjurjo, Pastorino and
  Naya}]{rodriguez_2021}
\bibinfo{author}{Rodr{\'\i}guez, A.J.}, \bibinfo{author}{Sanjurjo, E.},
  \bibinfo{author}{Pastorino, R.}, \bibinfo{author}{Naya, M.{\'A}.},
  \bibinfo{year}{2021}.
\newblock \bibinfo{title}{State, parameter and input observers based on
  multibody models and kalman filters for vehicle dynamics}.
\newblock \bibinfo{journal}{Mechanical Systems and Signal Processing}
  \bibinfo{volume}{155}, \bibinfo{pages}{107544}.
\bibitem[{Rozyn and Zhang(2010)}]{rozyn_2010}
\bibinfo{author}{Rozyn, M.}, \bibinfo{author}{Zhang, N.}, \bibinfo{year}{2010}.
\newblock \bibinfo{title}{A method for estimation of vehicle inertial
  parameters}.
\newblock \bibinfo{journal}{Vehicle system dynamics} \bibinfo{volume}{48},
  \bibinfo{pages}{547--565}.
\bibitem[{Savaresi et~al.(2010)Savaresi, Poussot-Vassal, Spelta, Sename and
  Dugard}]{savaresi_2010}
\bibinfo{author}{Savaresi, S.M.}, \bibinfo{author}{Poussot-Vassal, C.},
  \bibinfo{author}{Spelta, C.}, \bibinfo{author}{Sename, O.},
  \bibinfo{author}{Dugard, L.}, \bibinfo{year}{2010}.
\newblock \bibinfo{title}{Semi-active suspension control design for vehicles}.
\newblock \bibinfo{publisher}{Elsevier}.
\bibitem[{Solmaz et~al.(2008)Solmaz, Akar, Shorten and Kalkkuhl}]{selim_2008}
\bibinfo{author}{Solmaz, S.}, \bibinfo{author}{Akar, M.},
  \bibinfo{author}{Shorten, R.}, \bibinfo{author}{Kalkkuhl, J.},
  \bibinfo{year}{2008}.
\newblock \bibinfo{title}{Real-time multiple-model estimation of centre of
  gravity position in automotive vehicles}.
\newblock \bibinfo{journal}{Vehicle System Dynamics} \bibinfo{volume}{46},
  \bibinfo{pages}{763--788}.
\bibitem[{Tavernini et~al.(2020)Tavernini, Vacca, Metzler, Savitski, Ivanov,
  Gruber, Hartavi, Dhaens and Sorniotti}]{tavernini_2020}
\bibinfo{author}{Tavernini, D.}, \bibinfo{author}{Vacca, F.},
  \bibinfo{author}{Metzler, M.}, \bibinfo{author}{Savitski, D.},
  \bibinfo{author}{Ivanov, V.}, \bibinfo{author}{Gruber, P.},
  \bibinfo{author}{Hartavi, A.E.}, \bibinfo{author}{Dhaens, M.},
  \bibinfo{author}{Sorniotti, A.}, \bibinfo{year}{2020}.
\newblock \bibinfo{title}{An explicit nonlinear model predictive abs controller
  for electro-hydraulic braking systems}.
\newblock \bibinfo{journal}{IEEE Transactions on Industrial Electronics}
  \bibinfo{volume}{67}, \bibinfo{pages}{3990--4001}.
\newblock \DOIprefix\doi{10.1109/TIE.2019.2916387}.
\bibitem[{Theunissen et~al.(2020)Theunissen, Sorniotti, Gruber, Fallah, Ricco,
  Kvasnica and Dhaens}]{theunissen_2020}
\bibinfo{author}{Theunissen, J.}, \bibinfo{author}{Sorniotti, A.},
  \bibinfo{author}{Gruber, P.}, \bibinfo{author}{Fallah, S.},
  \bibinfo{author}{Ricco, M.}, \bibinfo{author}{Kvasnica, M.},
  \bibinfo{author}{Dhaens, M.}, \bibinfo{year}{2020}.
\newblock \bibinfo{title}{Regionless explicit model predictive control of
  active suspension systems with preview}.
\newblock \bibinfo{journal}{IEEE Transactions on Industrial Electronics}
  \bibinfo{volume}{67}, \bibinfo{pages}{4877--4888}.
\bibitem[{Thornton and Marion(2004)}]{thornton_2004}
\bibinfo{author}{Thornton, S.}, \bibinfo{author}{Marion, J.},
  \bibinfo{year}{2004}.
\newblock \bibinfo{title}{Classical Dynamics of Particles and Systems}.
\newblock \bibinfo{publisher}{Brooks/Cole}.
\bibitem[{VI-Grade(2022)}]{vigrade_2022}
\bibinfo{author}{VI-Grade}, \bibinfo{year}{2022}.
\newblock \bibinfo{title}{Vi-carrealtime}.
\newblock
  \bibinfo{howpublished}{\url{https://www.vi-grade.com/en/products/vi-carrealtime/}}.
\newblock \bibinfo{note}{[Online; accessed 23-March-2022]}.
\bibitem[{Viehweger et~al.(2021)Viehweger, Vaseur, van Aalst, Acosta, Regolin,
  Alatorre, Desmet, Naets, Ivanov, Ferrara et~al.}]{viehweger_2021}
\bibinfo{author}{Viehweger, M.}, \bibinfo{author}{Vaseur, C.},
  \bibinfo{author}{van Aalst, S.}, \bibinfo{author}{Acosta, M.},
  \bibinfo{author}{Regolin, E.}, \bibinfo{author}{Alatorre, A.},
  \bibinfo{author}{Desmet, W.}, \bibinfo{author}{Naets, F.},
  \bibinfo{author}{Ivanov, V.}, \bibinfo{author}{Ferrara, A.}, et~al.,
  \bibinfo{year}{2021}.
\newblock \bibinfo{title}{Vehicle state and tyre force estimation:
  demonstrations and guidelines}.
\newblock \bibinfo{journal}{Vehicle system dynamics} \bibinfo{volume}{59},
  \bibinfo{pages}{675--702}.
\bibitem[{Wang et~al.(2020)Wang, Wang, Zhang, Cao and Dorrell}]{wang_2020}
\bibinfo{author}{Wang, C.}, \bibinfo{author}{Wang, Z.}, \bibinfo{author}{Zhang,
  L.}, \bibinfo{author}{Cao, D.}, \bibinfo{author}{Dorrell, D.G.},
  \bibinfo{year}{2020}.
\newblock \bibinfo{title}{A vehicle rollover evaluation system based on
  enabling state and parameter estimation}.
\newblock \bibinfo{journal}{IEEE Transactions on Industrial Informatics}
  \bibinfo{volume}{17}, \bibinfo{pages}{4003--4013}.
\bibitem[{Wenzel et~al.(2006)Wenzel, Burnham, Blundell and
  Williams}]{wenzel_2006}
\bibinfo{author}{Wenzel, T.A.}, \bibinfo{author}{Burnham, K.},
  \bibinfo{author}{Blundell, M.}, \bibinfo{author}{Williams, R.},
  \bibinfo{year}{2006}.
\newblock \bibinfo{title}{Dual extended kalman filter for vehicle state and
  parameter estimation}.
\newblock \bibinfo{journal}{Vehicle system dynamics} \bibinfo{volume}{44},
  \bibinfo{pages}{153--171}.
\bibitem[{Wielitzka et~al.(2015)Wielitzka, Dagen and Ortmaier}]{wielitzka_2015}
\bibinfo{author}{Wielitzka, M.}, \bibinfo{author}{Dagen, M.},
  \bibinfo{author}{Ortmaier, T.}, \bibinfo{year}{2015}.
\newblock \bibinfo{title}{Joint unscented kalman filter for state and parameter
  estimation in vehicle dynamics}, in: \bibinfo{booktitle}{2015 IEEE Conference
  on Control Applications (CCA)}, \bibinfo{organization}{IEEE}. pp.
  \bibinfo{pages}{1945--1950}.
\bibitem[{Yang et~al.(2008)Yang, Liu and Cheng}]{yang_2008}
\bibinfo{author}{Yang, S.}, \bibinfo{author}{Liu, T.}, \bibinfo{author}{Cheng,
  Y.}, \bibinfo{year}{2008}.
\newblock \bibinfo{title}{Automatic measurement of payload for heavy vehicles
  using strain gages}.
\newblock \bibinfo{journal}{Measurement} \bibinfo{volume}{41},
  \bibinfo{pages}{491--502}.
\bibitem[{Zhu et~al.(2019)Zhu, Wang, Zhang and Zhang}]{zhu_2019}
\bibinfo{author}{Zhu, J.}, \bibinfo{author}{Wang, Z.}, \bibinfo{author}{Zhang,
  L.}, \bibinfo{author}{Zhang, W.}, \bibinfo{year}{2019}.
\newblock \bibinfo{title}{State and parameter estimation based on a modified
  particle filter for an in-wheel-motor-drive electric vehicle}.
\newblock \bibinfo{journal}{Mechanism and Machine Theory}
  \bibinfo{volume}{133}, \bibinfo{pages}{606--624}.

\end{thebibliography}

\end{document}